\documentclass[lettersize, journal]{IEEEtran}
\usepackage{float}
\usepackage{caption}

\usepackage{color}

\usepackage{cite}
\usepackage{amsmath,amssymb,amsfonts}
\usepackage{amsthm}
\usepackage{graphicx}
\usepackage{textcomp}
\usepackage{xcolor}
\usepackage{svg}
\usepackage{subcaption}

\usepackage{mathtools}

\usepackage[ruled,vlined]{algorithm2e}
\usepackage{algpseudocode}
\usepackage{epstopdf}

\usepackage{stfloats}
\newtheorem{definition}{Definition}
\newtheorem{remark}{Remark}
\newtheorem{theorem}{Theorem}

\newtheorem{lemma}{Lemma}

\newtheorem{corollary}{Corollary}

\allowdisplaybreaks
\makeatletter
\def\ScaleIfNeeded{%
\ifdim\Gin@nat@width>\linewidth \linewidth \else \Gin@nat@width
\fi } \makeatother

\graphicspath{{Fig/}}

\begin{document}

\title{Waveguide Division Multiple Access for Pinching-Antenna Systems (PASS)}
\author{Jingjing Zhao, Xidong Mu, Kaiquan Cai, Yanbo Zhu, and Yuanwei Liu,~\IEEEmembership{Fellow,~IEEE}
\thanks{J. Zhao, K. Cai, Y. Zhu are with the School of Electronics and Information Engineering, Beihang University, Beijing, China. (e-mail:\{jingjingzhao, ckq, zyb\}@buaa.edu.cn). X. Mu is with the Centre for Wireless Innovation (CWI), Queen's University Belfast, Belfast, BT3 9DT, U.K. (e-mail: x.mu@qub.ac.uk). Y. Liu is with the Department of Electrical and Electronic Engineering, the University of Hong Kong, Hong Kong, China (e-mail: yuanwei@hku.hk). }
}
\maketitle
\begin{abstract}
A novel concept of \textit{waveguide division multiple access (WDMA)} is proposed for multi-user pinching-antenna systems (PASS). The key principle of WDMA is to allocate each user with a dedicated waveguide, which is regarded as a new type of radio resources, so as to facilitate multi-user communications. By adjusting the activation positions of pinching antennas (PAs) over each waveguide, the \textit{pinching beamforming} can be exploited for intended user signal enhancement and inter-user interference mitigation. Considering both ideal continuous and practical discrete PA position activation schemes, a joint power allocation and pinching beamforming optimization problem is formulated for the maximization of the sum rate. An alternating optimization-based algorithm is developed to address the formulated non-convex problem. For solving the power allocation subproblem, the successive convex approximation method is invoked. For the pinching beamforming design subproblem, a penalty-based gradient ascent algorithm is first developed for the continuous PA activation case. Then, for the discrete PA activation case, a matching theory-based algorithm is proposed to achieve the near-optimal performance but with a low complexity. Numerical results unveil that: 1) For both continuous and discrete activation cases, PASS can achieve a significant performance gain over conventional fixed-position antenna systems; 2) the proposed WDMA can effectively underpin multi-user communications with the near orthogonality in free space achieved by the pinching beamforming; and 3) the performance gap between the discrete and continuous activation cases can be significantly alleviated with practically feasible numbers of PA candidate positions.
\end{abstract}
\section{Introduction}
With the proliferation of high-volume data transmission in recent years, extensive research efforts have been devoted to boost the wireless networks capacity. Among the candidate solutions, the multiple-input multiple-output (MIMO) technique has been regarded as a flagship one, which provides substantial antenna gain by exploiting the spatial-domain diversity or channel correlations~\cite{6736761,6375940}. In conventional MIMO communications, antennas are deployed with fixed locations either at the transmitter or receiver side, which can not fully leverage spatial degrees of freedom (DoFs) and thus leads to array gain loss within the confined antenna region. Upon recognizing this issue, researcher have started to explore the assets of flexible-antenna systems such as reconfigurable intelligent surfaces ~\cite{marco,8910627}, dynamic metasurface antennas~\cite{9324910}, holographic MIMO surfaces~\cite{9136592}, fluid antennas~\cite{9264694}, and movable antennas~\cite{10318061}.  Capitalizing on flexibly changing antennas position within a confined region, flexible-antenna systems can proactively construct favorable channel conditions and enhance the effective channel gain.

Despite the aforementioned benefits, existing flexible-antenna systems face certain limitations. On the one hand, higher-frequency bands are anticipated for expanded utilization, which suffers from severe spreading loss and thus poses significant difficulty to long-range transmission. For example, the signal transmission distance over the RIS-assisted link is generally longer than that over the direct link and suffers from the double-fading issue~\cite{marco}. Furthermore, high-frequency electromagnetic waves are also vulnerable to blockage. Since the moving range of antennas is restricted in the order of several to tens of wavelengths, existing fluid and movable antennas systems can only combat the small-scale fading, but cannot significantly alter the large-scale path loss given the limited operating space. 
On the other hand, current flexible-antenna systems have limited flexibility of antennas structure reconfiguration, i.e., adding or removing antennas on manufactured arrays is not easy to implement. 

\begin{figure}
\centerline{\includegraphics[width=1.0\linewidth]{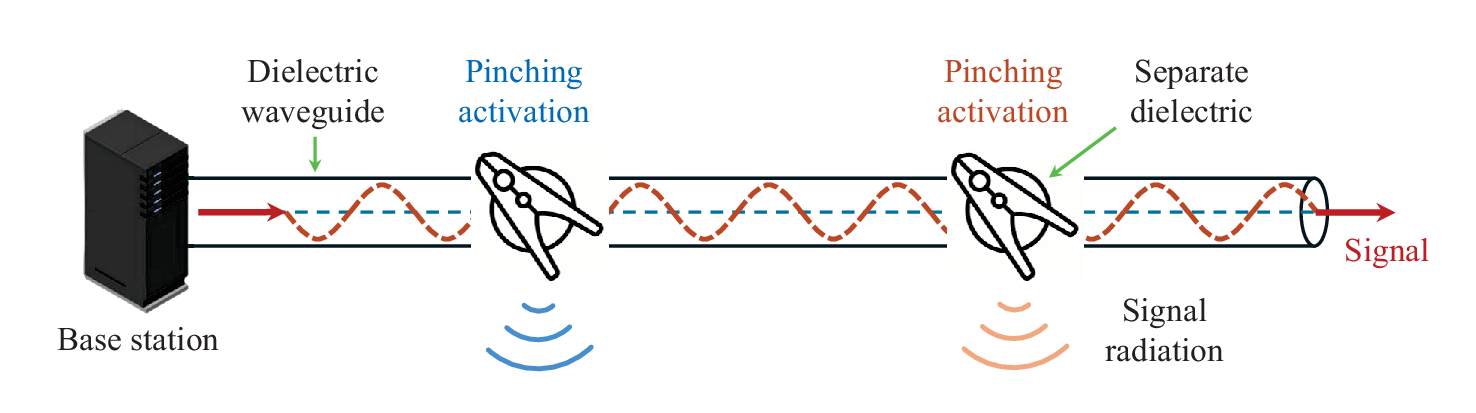}}
\caption{Illustration of pinching-antenna systems.}
\label{fig:PASS-model}
\end{figure}

The \textit{pinching-antenna system (PASS)} has been proposed as a groundbreaking technique in the family of flexible-antenna systems~\cite{docomo,ding,yuanwei}. The first prototype of PASS was demonstrated by NTT DOCOMO in 2022. As shown in Fig.~\ref{fig:PASS-model}, PASS employs a dielectric waveguide whose length spans from a few meters to tens of meters, where multiple radiation points can be activated by pinching separated dielectric particles.
These dielectric particles are referred to as \textit{pinching antennas (PAs)}, which allow part of the radio waves traveling through the waveguide to be induced into the free space. As such, flexible PA activation along the waveguide can facilitate the rapid establishment of diverse communication zones tailored to specific locations and environmental conditions.
Likewise, this radiation effect disappears upon releasing the pinch, allowing for instant termination of radio-wave emission and the elimination of the associated communication zone. Moreover, a PA can also function as a receiver, capturing external signals at its pinching point.

Considering the above mentioned distinctive characteristics of PASS, the main benefits of PASS compared to existing flexible-antenna systems are outlined as follows:
\begin{itemize}
    \item \textbf{Large-scale path loss mitigation}: Since the PA activation locations can be flexibly adjusted over a wide range, PASS can establish desired line-of-sight (LoS) links by surpassing obstacles. Moreover, given the negligible in-waveguide path loss, deploying PAs near to the target receiver can avoid the conventional long-range path loss via the facilitated ``last-meter" wireless transmission.
    \item \textbf{Scalable deployment}: The PASS configuration can be easily adjusted with the addition or removal of PAs along the waveguide. This design makes PASS more scalable than existing flexible-antenna systems for network deployment.  
    \item \textbf{Efficient pinching beamforming}: Each activated PA is able to induce a certain phase shift and amplitude change to the radiated radio wave. Collaboratively manipulating multiple PAs, the pinching beamforming can be achieved to strengthen desired signals and/or mitigate interference.
\end{itemize}

The above benefits have sparked initial research efforts towards PASS. In~\cite{ding}, the authors developed analytical results for PASS in three different cases, i.e., a single PA on a single waveguide, multiple PAs on a single waveguide, and multiple PAs on multiple waveguides. Simulation results were provided to demonstrate the superiority of PASS compared to conventional antenna systems. 
Since how to effectively activate PAs at proper locations is an essential problem in PASS, several studies have focused on the activation algorithms design~\cite{yanqing,kaidi, ali, guojia}. Considering the rate maximization problem for downlink PASS-enabled communications, the authors of~\cite{yanqing} proposed a two-stage activation algorithm with low complexity and near-optimal performance, where the large-scale path loss minimization and the constructive signals combination were addressed separately. Incorporating the NOMA technique, the multi-user communications scenario with a single waveguide was considered in~\cite{kaidi}, where the matching algorithm was developed for determining what locations and how many PAs to be activated for the sum rate maximization. In~\cite{ali}, the hybrid beamforming problem was investigated, where the transmit and pinching beamformimg were iteratively updated to maximize the weighted sum rate. Moreover, the machine learning was leveraged in~\cite{guojia} for the joint transmit and pinching beamforming in PASS, which achieved superior performance than a heuristic baseline with low inference complexity.

\subsection{Motivations and Contributions}
\begin{figure}
\centerline{\includegraphics[width=1.0\linewidth]{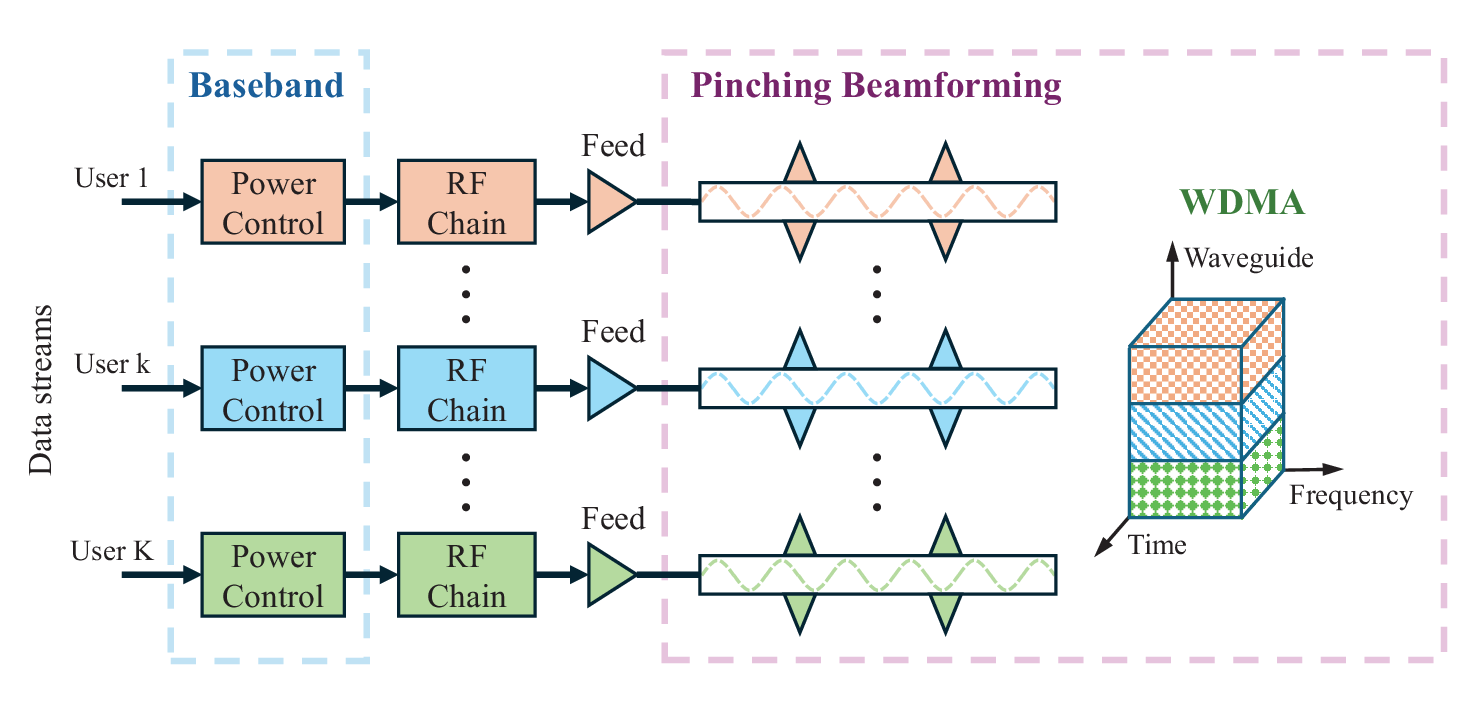}}
\caption{Proposed PASS-enabled multi-user communications with the WDMA.}
\label{fig:transmission-architecture}
\end{figure}
Note that one of the key features of PASS is that each waveguide can only convey the same data stream to PAs, which is different from conventional MIMO systems~\cite{ding}. As a result, the mutli-user communication design is a vital issue for unlocking the PASS potentials. The existing multi-user communication studies in PASS~\cite{ali,guojia} focused on the joint transmit and pinching beamforming, where the signal sent through each waveguide is a superimposed mixture of multiple users' signals and conventional baseband transmit beamforming is exploited for spatial multiplexing. This, however, requires the sophisticated basedband signal processing and high computational complexity caused by the highly coupling between transmit and pinching beamforming designs. Moreover, considering the practical deployment of PASS, multiple waveguides are likely to be deployed in geographically separate areas~\cite{docomo}. For example, different waveguides can be installed on the ceiling in different rooms for serving different local in-door users. In this case, transmitting superimposed signals over each waveguide is not either essential or efficient. An intriguing question is then put forward: \textit{Is there an alternative and simple option for employing PASS into multi-user communications?}


Motivated by this question, we propose a simple design for facilitating PASS-enabled multi-user communications with the novel concept of \textit{waveguide division multiple access (WDMA)}. As illustrated in Fig.~\ref{fig:transmission-architecture}, the key idea of WDMA is to assign each user with a dedicated waveguide, which is regarded as a new type of radio resources similar to time/frequency resource blocks. The pinching beamforming on each waveguide is then exploited for desired signal enhancement and inter-user interference mitigation with the aim of achieving the (near) orthogonality in the free space. By doing so, each waveguide only has to convey a single data stream for its intended user, and therefore the baseband essentially carries out the simple power allocation, which significantly reduces the design complexity.

To examine the effectiveness of the proposed WDMA, we investigate the resultant joint power allocation and pinching beamforming problem in PASS-enabled multi-user communications. The main contributions of this paper are summarized as follows:
\begin{itemize}
    \item We propose a simple multi-user communications design for PASS with the novel concept of WDMA, where each user is served by one dedicated allocated waveguide and pinching beamforming is used to facilitate the (near) orthogonality in the free-space transmission. Considering both ideal continuous and practical discrete PA position activation, a joint power allocation and pinching beamforming optimization problem is formulated for maximizing the sum rate, while satisfying the minimum rate requirements of users and positions constraints of PAs.
    \item We develop an alternative optimization (AO)-based framework for solving the resulting non-convex problem in both continuous and discrete activation cases, which decouples the original problem into two subproblems. Specifically, for given pinching beamforming, the successive convex approximation (SCA) method is advocated for solving the power allocation subproblem. 
    For the pinching beamforming design with given power allocation, we first develop a penalty-based gradient ascent algorithm (GAA) for the continuous activation case. Then, a matching theory-based algorithm is developed to obtain the near-optimal pinching beamforming for the discrete activation case, while maintaining low computational complexity.    
    
    \item Numerical results unveil that 1) For both ideal and practical cases, PASS achieves superior performance than conventional antenna systems; 2) the WDMA can support multi-user communications with the near orthogonality in free space achieved by the pinching beamforming; and 3) the performance gap between the discrete and continuous activation cases can be alleviated with a practical number of PA candidate positions. 
\end{itemize}

\subsection{Organization and Notations}
The rest of this paper is structured as follows. In Section II, the proposed PASS-enabled multi-user communications with the WDMA is first introduced, which is followed by the sum rate maximization problem formulation. In Section III, the AO-based optimization framework is developed to address the resulting non-convex optimization problem for both continuous and discrete activation cases. Section IV presents the simulation results, and Section V concludes the paper. 

$\textit {Notations}$: Scalars, vectors, and matrices are denoted by italic letters, bold-face lower-case, and bold-face upper-case, respectively. $\mathbb{C}^{N\times M}$ denotes the set of $N\times M$ complex-valued matrices. Superscripts $(\cdot)^*, (\cdot)^T, (\cdot)^H$, and $(\cdot)^{-1}$ denote the conjugate, transpose, conjugate transpose, and inversion operators, respectively. $|\cdot|$ and  $\left\|\cdot\right\|$ denote the  determinant and Euclidean norm of a matrix, respectively. $\text{Tr}\left(\cdot\right)$, $\left\|\cdot\right\|_F$, and $\text{vec}\left(\cdot\right)$ denote the trace, Frobenius norm, and vectorization of a matrix,  respectively. $[\cdot]_{m,n}$ denotes the $(m,n)$-th element of a matrix. $\mathbf{1}_{{N}}$ denotes the all-one row vector with length $N$. 
$\mathbb{E}$ denotes the expectation operator. $\circ$ denotes the Hadamard multiplication. All random variables are assumed to be \textit{zero} mean. 
\section{System Model and Problem Formulation}
In this section, we present the system model of the PASS-enabled multi-user communications with the WDMA, and formulate the joint power allocation and pinching beamforming problem for maximizing the users' sum rate. 
\subsection{System Description}
\begin{figure}
\centerline{\includegraphics[width=0.9\linewidth]{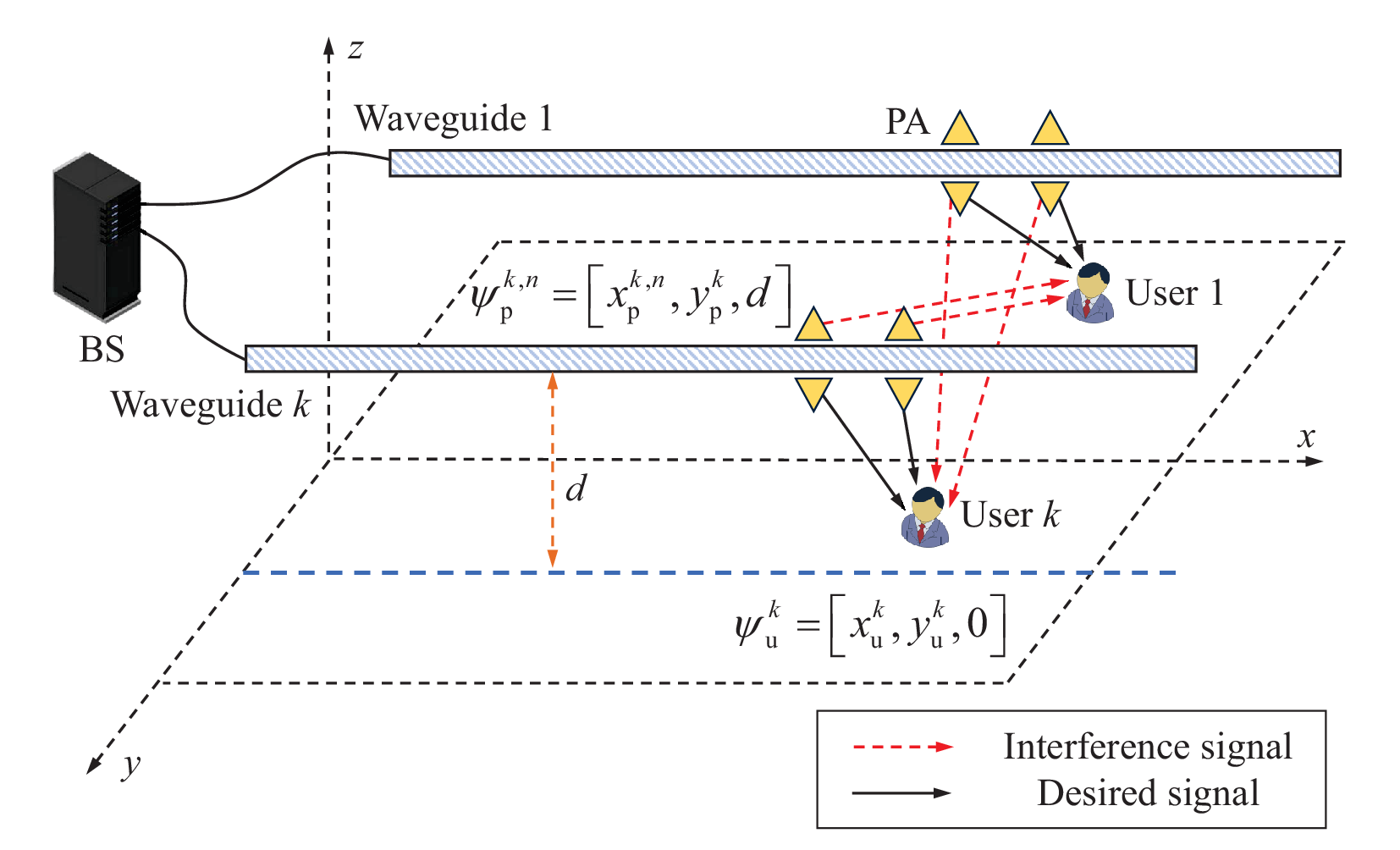}}
\caption{Illustration of the PASS-enabled multi-user communication system.}
\label{fig:system_model}
\end{figure}
The system model of the proposed PASS-enabled multi-user communications is shown in Fig.~\ref{fig:system_model}, where the BS consisting of $K$ dielectric waveguides serves $K$ users via the WDMA. Specifically, the BS baseband essentially carries out power allocation to each user's data stream. Each waveguide with $N$ PAs conveys one data stream and serves the intended user via pinching beamforming. 
For ease of notation, we use the same index $k$ for each waveguide and its serving user, and denote the set of users and waveguides as $\mathcal{K}_{\text{u}}$ and $\mathcal{K}_{\text{w}}$, respectively. The location of the $k$-th user is denoted by $\boldsymbol{\psi}_{\text{u}}^k=\left[x_{\text{u}}^k,y_{\text{u}}^k,0\right]$. Without loss of generality, we assume that all the waveguides are aligned parallel to the $x$-axis with equal height of $d$. Then, the location of the $n$-th antenna on the $k$-th waveguide is denoted by $\boldsymbol{\psi}_{\text{p}}^{k,n} = \left[x_{\text{p}}^{k,n}, y_{\text{p}}^{k}, d\right]$, and the set of $N$ PAs $x$-axis locations on the $k$-th waveguide is represented by $\mathbf{x}_{\text{p}}^{k} = \left[x_{\text{p}}^{k,1}, ..., x_{\text{p}}^{k,N}\right]$.  Let $\bar{\boldsymbol{\psi}}^k_{\text{p}}=[0,y_{\text{p}}^k,d]$ denote the position of the feed point of the $k$-th waveguide. Assume that PAs on each waveguide are placed in a successive order, i.e., $x_{\text{p}}^{k,{n+1}}>x_{\text{p}}^{k,{n}}, \forall 1\leq n< N, \forall k$, and the maximum deployment range of PAs is $L$. 
We consider both the ideal and practical cases for PA activation as follows:
\begin{itemize}
    \item \textbf{Ideal case - continuous activation}: In this case, PAs can be flexibly activated on each waveguide over the distance significantly larger than the wavelength, as long as the minimum spacing is guaranteed to avoid antenna coupling. Thus, the feasible set of $\mathbf{x}_{\text{p}}^k$ can be expressed as
    \begin{equation}
        \mathcal{F}^{\text{C}} = \left\{\left.\mathbf{x}_{\text{p}}^k\right|0\leq x_{\text{p}}^{k,{n}}\leq L, x_{\text{p}}^{k,{n+1}} - x_{\text{p}}^{k,{n}}\geq \Delta\right\},
    \end{equation}
    where $\Delta$ denotes the minimum spacing among PAs. 
    \item \textbf{Practical case - discrete activation}: In this case, PAs can only be activated at preconfigured locations. Denote the set of $A$ available locations on each waveguide as $\mathcal{S}_{k} = \left\{\bar{x}_{\text{p}}^{k,1}, ..., \bar{x}_{\text{p}}^{k,A}\right\}$ with $A\geq N$. We assume that positions are uniformly spaced along the waveguide, and thus the spacing between any two adjacent positions is $L/(A-1)$.
    Then the feasible set of $\mathbf{x}_{\text{p}}^k$ can be expressed as
    \begin{equation}
        \mathcal{F}^{\text{D}} = \left\{\left.\mathbf{x}_{\text{p}}^k\right|{x}_{\text{p}}^{k,n}\in \mathcal{S}_{k}, {x}_{\text{p}}^{k,n}\neq {x}_{\text{p}}^{k,n+1}\right\}.
    \end{equation}
\end{itemize}
\subsection{Channel and Signal Model}
The wireless channel between the PAs on the $k$-th waveguide and the user $k', \forall k'\in\mathcal{K}_{\text{u}}$ is given by
\begin{align}
    & \mathbf{h}_{k,k'}\left(\mathbf{x}_{\text{p}}^{k}\right)\nonumber\\
    &= \left[\frac{\eta e^{-j\frac{2\pi}{\lambda}\left\|\boldsymbol{\psi}^{k'}_{\text{u}}-\boldsymbol{\psi}^{k,1}_{\text{p}}\right\|}}{\left\|\boldsymbol{\psi}^{k'}_{\text{u}}-\boldsymbol{\psi}^{k,1}_{\text{p}}\right\|}, ..., \frac{\eta e^{-j\frac{2\pi}{\lambda}\left\|\boldsymbol{\psi}^{k'}_{\text{u}}-\boldsymbol{\psi}_{\text{p}}^{k,N}\right\|}}{\left\|\boldsymbol{\psi}^{k'}_{\text{u}}-\boldsymbol{\psi}_{\text{p}}^{k,N}\right\|}\right]^T,
\end{align}
where $\eta=\frac{\lambda}{4\pi}$ represents the channel gain at the reference distance of $1$~m, $\lambda$ denotes the signal wavelength in the free space, and $\left\|\boldsymbol{\psi}^{k'}_{\text{u}}-\boldsymbol{\psi}_{\text{p}}^{k,n}\right\|$ is given by
\begin{equation}
   \left\|\boldsymbol{\psi}^{k'}_{\text{u}}-\boldsymbol{\psi}_{\text{p}}^{k,n}\right\|=\sqrt{\left(x_{\text{u}}^{k'}-x_{\text{p}}^{k,n}\right)^2+\left(y_{\text{u}}^{k'}-y_{\text{p}}^{k}\right)^2+d^2}. 
\end{equation}

Let $\mathbf{g}_k\in\mathbb{C}^{N\times 1}$ denote the propagation channel on the $k$-th waveguide, which is given by
\begin{align}
\label{eq:waveguide-channel}
    \mathbf{g}_{k}\left(\mathbf{x}_{\text{p}}^{k}\right) & = \left[e^{-j\frac{2\pi \left\|\boldsymbol{\psi}_{\text{p}}^{k,1}-\bar{\boldsymbol{\psi}}_{\text{p}}^k\right\|}{\lambda_{\text{g}}}},..., e^{-j\frac{2\pi \left\|\boldsymbol{\psi}_{\text{p}}^{k,N}-\bar{\boldsymbol{\psi}}_{\text{p}}^k\right\|}{\lambda_{\text{g}}}}\right]^T\nonumber\\
    &=\left[e^{-j\frac{2\pi x_{\text{p}}^{k,1}}{\lambda_{\text{g}}}},..., e^{-j\frac{2\pi x_{\text{p}}^{k,N}}{\lambda_{\text{g}}}}\right]^T,
\end{align}
where $\lambda_{\text{g}}=\frac{\lambda}{n_{\text{eff}}}$ denotes the guided wavelength with $n_{\text{eff}}$ representing the effective refractive index of a dielectric waveguide~\cite{microwave}. Note that the waveguide propagation loss is omitted in~\eqref{eq:waveguide-channel}, which is negligible compared to the free space path loss as demonstrated in~\cite{kaidi}. Let $s_k$ denote the normalized signal transmitted through the $k$-th waveguide with $\mathbb{E}\left[{s}_k{s}_k^*\right] = 1$,
then the signal received by the $k$-th user is given as follows:
\begin{align}
        y_k = &\underbrace{\mathbf{h}^T_{k,k}\left(\mathbf{x}_{\text{p}}^{k}\right)\mathbf{g}_{k}\left(\mathbf{x}_{\text{p}}^{k}\right)\sqrt{\frac{p_k}{N}}s_k}_{\text{desired signal}} \nonumber\\
    & + \underbrace{\sum_{k'\neq k}^K \mathbf{h}^T_{k',k}\left(\mathbf{x}_{\text{p}}^{k'}\right)\mathbf{g}_{k'}\left(\mathbf{x}_{\text{p}}^{k'}\right)\sqrt{\frac{p_{k'}}{N}}s_{k'}}_{\text{interfering signal}} + n_k,
\end{align}
where $p_k$ denotes the power allocated to the $k$-th user, and $n_k\sim \mathcal{CN}(0, \sigma_k^2)$ represents the additive white Gaussian noise (AWGN) at the $k$-th user, with $\sigma_k^2$ denoting the noise power. Assume that the total power on each waveguide is equally distributed among $N$ PAs, which leads to the per-antenna power of $\frac{p_k}{N}$. Then, the achievable rate at user $k$ can be expressed as
\begin{align}
    & R_k\nonumber\\
    &= \log _2\left(1+\frac{p_k\left|\mathbf{h}^T_{k,k}\left(\mathbf{x}_{\text{p}}^{k}\right)\mathbf{g}_{k}\left(\mathbf{x}_{\text{p}}^{k}\right)\right|^2}{\sum\limits_{k'\neq k}^{K}p_{k'}\left|\mathbf{h}^T_{k',k}\left(\mathbf{x}_{\text{p}}^{k'}\right)\mathbf{g}_{k'}\left(\mathbf{x}_{\text{p}}^{k'}\right)\right|^2+N\sigma_k^2}\right)\nonumber\\
    &= \log_2\left(1+\frac{p_k\left|\sum\limits_{n=1}^N\frac{\eta e^{-j\phi_{k,k}^n}}{\left\|\boldsymbol{\psi}_{\text{u}}^k-\boldsymbol{\psi}_{\text{p}}^{k,n}\right\|}\right|^2}{\sum\limits_{k'\neq k}^Kp_{k'}\left|\sum\limits_{n=1}^N\frac{\eta e^{-j\phi_{k',k}^n}}{\left\|\boldsymbol{\psi}_{\text{u}}^{k}-\boldsymbol{\psi}_{\text{p}}^{k',n}\right\|}\right|^2+N\sigma_k^2}\right),
    \label{eq:data-rate}
\end{align}
where $\phi_{k,k}^n$ and $\phi_{k',k}^n$ are defined as
\begin{equation}
    \phi_{k,k}^n \triangleq \frac{2\pi}{\lambda}\left\|\boldsymbol{\psi}_{\text{u}}^k-\boldsymbol{\psi}_{\text{p}}^{k,n}\right\| + \frac{2\pi}{\lambda_g}\left\|\boldsymbol{\psi}_{\text{p}}^{k,n}-\bar{\boldsymbol{\psi}}_{\text{p}}^{k}\right\|,
\end{equation}
and 
\begin{equation}
    \phi_{k',k}^n \triangleq \frac{2\pi}{\lambda}\left\|\boldsymbol{\psi}_{\text{u}}^k-\boldsymbol{\psi}_{\text{p}}^{k',n}\right\| + \frac{2\pi}{\lambda_g}\left\|\boldsymbol{\psi}_{\text{p}}^{k',n}-\bar{\boldsymbol{\psi}}_{\text{p}}^{k'
    }\right\|,
\end{equation}
respectively.
\subsection{Problem Formulation}
We aim at maximizing the sum rate of all users by jointly optimizing the power allocation and the pinching beamforming, subject to the minimum rate requirements of users and position constraints of PAs. Let $\mathbf{X} = \left[\left(\mathbf{x}_{\text{p}}^{1}\right)^T, ..., \left(\mathbf{x}_{\text{p}}^{K}\right)^T\right]^T\in\mathbb{R}^{K\times N}$ and $\mathbf{p}=\left[p_1, ..., p_{K}\right]\in\mathbb{R}^{K}$.The optimization problem is formulated as follows: 
\begin{subequations}
\label{eq:optimization_problem}
\begin{equation}
\label{eq:objective-function}
(\text{P1}):	\max_{\mathbf{X},\mathbf{p}}\sum_{k=1}^{K}R_k,
\end{equation}
\begin{equation}
\label{eq:minimum-rate}
    {\rm{s.t.}} \ \ R_k\geq \bar{R}_k, \forall k,
\end{equation}
\begin{equation}
\label{eq:position-feasible-set}
    \mathbf{x}_{\text{p}}^k \in \mathcal{F}^{\text{C}}\slash \mathbf{x}_{\text{p}}^k \in \mathcal{F}^{\text{D}},
\end{equation}
\begin{equation}
    \label{eq:maximum-power}
 \sum_{k=1}^Kp_k\leq P_{\text{max}}, 
\end{equation}
\end{subequations}
where constraint~\eqref{eq:minimum-rate} ensures the minimum rate constraint of each user with required communication rate $\bar{R}_k$. Constraint~\eqref{eq:position-feasible-set} characterizes the corresponding feasible set of activation locations for the continuous or discrete case.  Constraint~\eqref{eq:maximum-power} restricts the BS maximum transmit power not to exceed $P_{\text{max}}$. 

Problem~\eqref{eq:optimization_problem} is intractable to solve due to the following two main reasons. First, the objective function~\eqref{eq:objective-function} and constraint~\eqref{eq:minimum-rate} are highly non-convex w.r.t. $\mathbf{X}$ and $\mathbf{p}$, and the involved variables are coupled. Second, for the discrete activation case, the problem becomes a mixed integer one that is challenging to search for the optimal solution due to the large number of discrete available locations. Generally, there is no established approach for efficiently solving such non-convex optimization problem. In the following, we propose an optimization framework by applying the AO method~\cite{7366709} for both continuous and discrete activation cases.    

\section{Proposed Joint Power Allocation and Pinching Beamforming Solution}
\begin{figure}
\centerline{\includegraphics[width=0.9\linewidth]{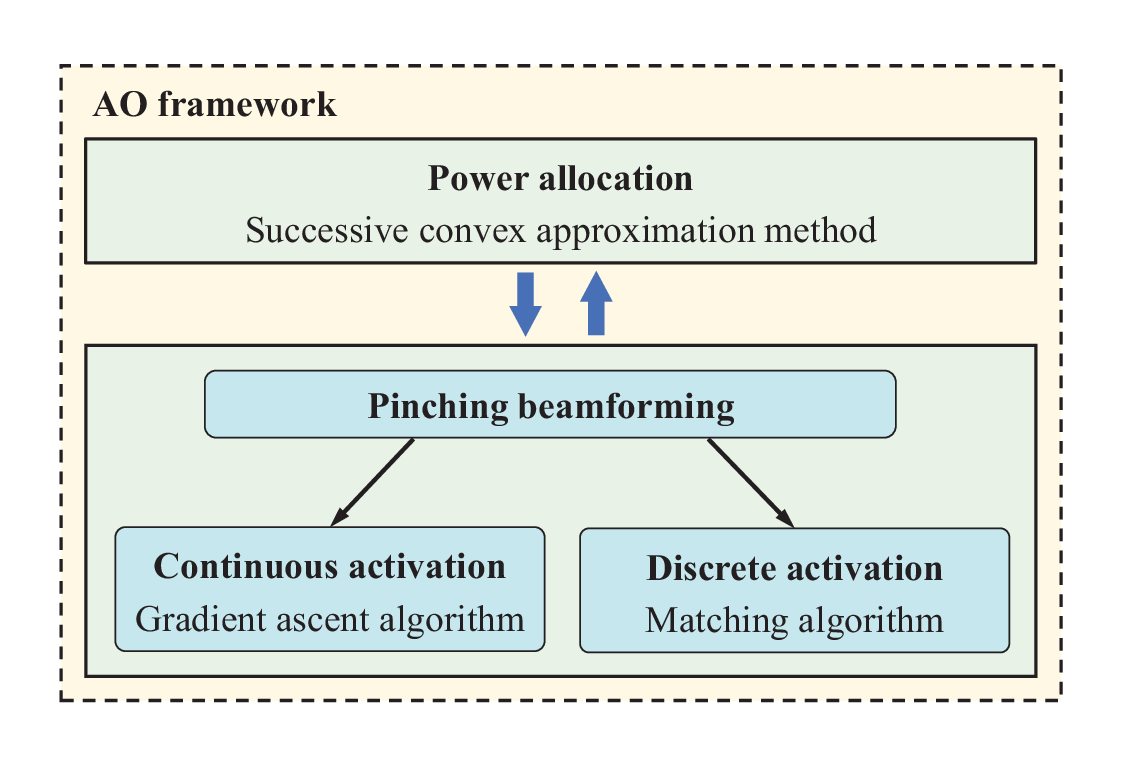}}
\caption{The proposed AO-based optimization framework.}
\label{fig:AO-optimization}
\end{figure}

The proposed AO-based optimization framework decouples the original problem~\eqref{eq:optimization_problem} into two subproblems, i.e., power allocation and pinching beamforming.
As shown in Fig.~\ref{fig:AO-optimization}, each group of variables is iteratively optimized while keeping the others fixed, resulting in an iterative optimization process. Specifically, for given pinching beamforming $\mathbf{X}$, we optimize the power allocation $\mathbf{p}$ based on the SCA method. For any given power allocation $\mathbf{p}$, the pinching beamforming $\mathbf{X}$ is optimized by applying the GAA~\cite{GDA} and the matching-theory based approach~\cite{matching1,matching2}, for the continuous and discrete activation cases, respectively. Finally, we present the overall algorithm design.

\subsection{Power Allocation}
For any given pinching beamforming $\mathbf{X}$, the power allocation variables $\mathbf{p}$ can be optimized by solving the following problem:
\begin{subequations}
\label{eq:power-optimization-problem}
\begin{equation}
\label{eq:power-objective-function}
(\text{P1-1}):	\max_{\mathbf{p}}\sum_{k=1}^{K}R_k,
\end{equation}
\begin{equation}
    {\rm{s.t.}} \ \ \eqref{eq:minimum-rate}, \eqref{eq:maximum-power}.
\end{equation}
\end{subequations}
Problem (P1-1) is a non-convex optimization problem due to the non-convex objective function~\eqref{eq:power-objective-function} and constraint~\eqref{eq:minimum-rate} w.r.t. $\mathbf{p}$.
Note that $R_k$ can be rewritten as:
\begin{equation}
    \label{eq:R_k-rewrite}
    R_k=\log_2\left(\sum_{k'=1}^Kp_{k'}\left|\mathbf{h}^T_{k',k}\mathbf{g}_{k'}\right|^2+N\sigma_k^2\right) - \check{R}_k,
\end{equation}
where 
\begin{equation}
    \check{R}_k = \log_2\left(\sum\limits_{k'\neq k}^{K}p_{k'}\left|\mathbf{h}^T_{k',k}\mathbf{g}_{k'}\right|^2+N\sigma_k^2\right).
\end{equation}
\eqref{eq:R_k-rewrite} is in the form of difference of two concave functions w.r.t. $\mathbf{p}$. As such, we can apply the SCA method to approximate $R_k$ with a concave lower bound in the $t$-th iteration as follows:
\begin{equation}
    R_k \geq \log_2\left(\sum_{k'=1}^Kp_{k'}\left|\mathbf{h}^T_{k',k}\mathbf{g}_{k'}\right|^2+N\sigma_k^2\right) - \check{R}_k^{\text{up}} \triangleq R_k^{\text{lw}},
\end{equation}
where $\check{R}_k^{\text{up}}$ is the first-order Taylor expansion of $\check{R}_k$ at the given local point $\mathbf{p}^{(t)}=\left[p_{1}^{(t)}, ..., p_K^{(t)}\right]$. $\check{R}_k^{\text{up}}$ is given as
\begin{align}
    & \check{R}_k^{\text{up}} =  \log_2\left(\sum\limits_{k'\neq k}^{K}p_{k'}^{(t)}\left|\mathbf{h}^T_{k',k}\mathbf{g}_{k'}\right|^2+N\sigma_k^2\right)\nonumber\\
    & + \sum_{k'\neq k}^{K}\frac{\left|\mathbf{h}_{k',k}^T\mathbf{g}_{k'}\right|^2\log_2(e)}{\sum\limits_{m\neq k}^Kp_{m}^{(t)}\left|\mathbf{h}_{m,k}^T\mathbf{g}_m\right|^2+N\sigma_k^2}\left(p_{k'}-p_{k'}^{(t)}\right).
\end{align}
As a result, for given point $\mathbf{p}$, by replacing $R_k$ with its concave lower bound, problem (P1-1) can be transformed into the following optimization problem:
\begin{subequations}
\label{eq:power-convex-problem}
\begin{equation}
\label{eq:power-convex-objective-function}
(\text{P1-1-1}):	\max_{\mathbf{p}}\sum_{k=1}^{K}R_k^{\text{lw}},
\end{equation}
\begin{equation}
    {\rm{s.t.}} \ \ R_k^{\text{lw}}\geq \bar{R}_k, \forall k,
\end{equation}
\begin{equation}
    \eqref{eq:maximum-power}.
\end{equation}
\end{subequations}
Problem (P1-1-1) is a convex optimization problem, which can be solved by standard solvers such as CVX~\cite{boyd2004convex}. The details of the developed power allocation algorithm is shown in~\textbf{Algorithm~\ref{alg:SCA}}. 
\begin{algorithm}[tp]
	\caption{Algorithm For Solving Problem (P1-1)}
        \label{alg:SCA}
	\LinesNumbered
	\KwIn{$\mathbf{X}$}
	Initialize $\mathbf{p}^{(0)}$;\\
    Set iteration index $t=0$;\\
    \Repeat{Increment of the sum rate is below $\epsilon$}{
    For given $\mathbf{p}^{(t)}$, solve the relaxed problem (P1-1-1);\\
    Update $\mathbf{p}^{(t+1)}$ with the obtained solution;\\
    $t=t+1$;
    }
    \KwOut{$\mathbf{p}$}
\end{algorithm}

Since the objective function value in~\eqref{eq:power-convex-objective-function} is non-decreasing after each SCA iteration and the sum rate is upper bounded due to the limited power consumption, the SCA algorithm is guaranteed to converge to a stationary point of the original problem (P1-1). For solving problem (P1-1-1), the computational complexity is estimated as $\mathcal{O}\left(K^{3.5}\right)$ if the interior point method is employed~\cite{local-convergence}. Then, the overall complexity of the SCA-based power allocation algorithm is $\mathcal{O}\left(I_{\text{pa}}K^{3.5}\right)$, where $I_{\text{pa}}$ is the total number of SCA iterations. 
\subsection{Pinching Beamforming with Continuous Activation}
Under the ideal assumption of continuous PA activation, with the power allocation $p_k, \forall k$ fixed, problem \eqref{eq:optimization_problem} can be reformulated as
\begin{subequations}
\label{eq:location_optimization_problem}
\begin{equation}
\label{eq:location_objective-function}
(\text{P1-2}):	\max_{\mathbf{X}}\sum_{k=1}^{K}R_k,
\end{equation}
\begin{equation}
    {\rm{s.t.}} \ \ \eqref{eq:minimum-rate},
\end{equation}
\begin{equation}
    0\leq x_{\text{p}}^{k,{n}}\leq L, \forall k,n,
    \label{eq:antenna-position-region}
\end{equation}
\begin{equation}
    \label{eq:antenna-distance}
    x_{\text{p}}^{k,{n+1}} - x_{\text{p}}^{k,{n}}\geq \Delta, \forall k, 1\leq n<N.
\end{equation}
\end{subequations}
As observed in~\eqref{eq:data-rate},  $R_k$ is highly complex and non-convex w.r.t. $\mathbf{X}$, as $\mathbf{X}$ affect both the numerators and denominators of the signal-to-noise ratios (SNRs) and also appear in the exponent of the complex-valued numbers. Therefore, it is challenging to obtain the global optimal solution for problem~(P1-2). {Next, we propose the GAA framework~\cite{GDA} for optimizing $\mathbf{X}$. }

Since the GAA can not solve constrained optimization problems directly, we need to firstly convert problem~(P1-2) to an unconstrained one via the penalty method~\cite{penalty-2}. Specifically, for constraint~\eqref{eq:antenna-position-region}, we introduce the auxiliary variables $\tilde{\mathbf{x}}^k_{\text{p}}\in\mathbb{R}^{N}=\left[\tilde{x}_{\text{p}}^{k,{1}}, ..., \tilde{x}_{\text{p}}^{k,{N}}\right], \forall k$ that satisfies
\begin{equation}
    \label{eq:auxiliaray-x}
    {\mathbf{x}}^k_{\text{p}} = \frac{L}{2}\left(1+\tanh \left(\tilde{\mathbf{x}}^k_{\text{p}}\right)\right), \forall k,
\end{equation}
where $\tanh \left(\tilde{\mathbf{x}}^k_{\text{p}}\right)$ is given by
\begin{equation}
    \tanh \left(\tilde{\mathbf{x}}^k_{\text{p}}\right)[a] = \frac{e^{\tilde{\mathbf{x}}^k_{\text{p}}[a]}-e^{-\tilde{\mathbf{x}}^k_{\text{p}}[a]}}{e^{\tilde{\mathbf{x}}^k_{\text{p}}[a]}+e^{-\tilde{\mathbf{x}}^k_{\text{p}}[a]}}\in\left(-1,1\right).
\end{equation}
The constraint~\eqref{eq:antenna-position-region} can be removed from problem~(P1-2) by replacing the optimization variables $\mathbf{x}_{\text{p}}^k, \forall k$ with $\tilde{\mathbf{x}}_{\text{p}}^k, \forall k$ based on~\eqref{eq:auxiliaray-x}. Let $\tilde{\mathbf{X}}=\left[\left(\tilde{\mathbf{x}}_{\text{p}}^{1}\right)^T, ..., \left(\tilde{\mathbf{x}}_{\text{p}}^{K}\right)^T\right]\in\mathbb{R}^{N\times K}$. Problem~(P1-2) can be transformed to 
\begin{subequations}
\label{eq:location_optimization_problem_1}
\begin{equation}
\label{eq:location_objective-function_1}
(\text{P1-2-1}):	\max_{\tilde{\mathbf{X}}}f\left(\tilde{\mathbf{X}}\right)\triangleq\sum_{k=1}^{K}R_k,
\end{equation}
\begin{equation}
\label{eq:minimum-rate_subproblem_1}
    {\rm{s.t.}} \ \ \ \ \zeta_{1,k}\left(\tilde{\mathbf{X}}\right)\triangleq \bar{R}_k-R_{k}\leq 0, \forall k,
\end{equation}
\begin{align}
\label{eq:antenna-distance_subproblem_1}
&\zeta_2\left(\tilde{x}_{\text{p}}^{k,{n+1}},\tilde{x}_{\text{p}}^{k,{n}}\right)\nonumber\\
&\triangleq \frac{2\Delta}{L}- \left(\tanh\left(\tilde{x}_{\text{p}}^{k,{n+1}}\right)-\tanh\left(\tilde{x}_{\text{p}}^{k,{n}}\right)\right) \leq 0, \nonumber\\
& \ \ \ \ \ \ \ \ \ \ \ \ \ \ \ \ \ \ \ \ \ \ \ \ \ \ \ \ \forall k, 1\leq n< N. 
\end{align}
\end{subequations}

For the remaining constraints~\eqref{eq:minimum-rate} and~\eqref{eq:antenna-distance}, we first transform them equivalently to the following equality constraints:
\begin{equation}
\label{eq:minimum-rate-equality}
    \kappa_{1,k}\left(\tilde{\mathbf{X}}\right)\triangleq\max\left\{0, \zeta_{1,k}\left(\tilde{\mathbf{X}}\right)\right\}=0, \forall k
\end{equation}
\begin{align}
    \label{eq:antenna-distance-equality}
    \kappa_2\left(\tilde{x}_{\text{p}}^{k,{n+1}},\tilde{x}_{\text{p}}^{k,{n}}\right)\triangleq&\max\left\{0, \zeta_2\left(\tilde{x}_{\text{p}}^{k,{n+1}},\tilde{x}_{\text{p}}^{k,{n}}\right)\right\}=0, \nonumber\\
    &\ \ \ \ \ \ \ \ \ \ \ \ \ \forall k, 1\leq n< N. 
\end{align}
Applying the penalty method, \eqref{eq:minimum-rate-equality} and~\eqref{eq:antenna-distance-equality} can be added to the objective function~\eqref{eq:location_objective-function_1} for penalizing the sum rate when constraints~\eqref{eq:minimum-rate_subproblem_1} and~\eqref{eq:antenna-distance_subproblem_1} are not satisfied. As such, problem (P1-2-1) can be transformed to the following unconstrained one:
\begin{equation}
        \label{eq:location_objective-function_2}
	(\text{P1-2-2}): \max_{\tilde{\mathbf{X}}}f\left(\tilde{{\mathbf{X}}}\right)-\beta\kappa_3\left(\tilde{{\mathbf{X}}}\right),
\end{equation}
{where}
\begin{equation}
    \kappa_3\left(\tilde{\mathbf{X}}\right)\triangleq\left(\sum\limits_{k=1}^{K}\kappa_{1,k}\left(\tilde{\mathbf{X}}\right) + \sum_{k=1}^{K}\sum\limits_{n=1}^{N-1}\kappa_2\left(\tilde{x}_{\text{p}}^{k,{n+1}},\tilde{x}_{\text{p}}^{k,{n}}\right)\right).
\end{equation}
Since $\kappa_{1,k}\left(\tilde{\mathbf{X}}\right)$ and $\kappa_2\left(\tilde{x}_{\text{p}}^{k,{n+1}},\tilde{x}_{\text{p}}^{k,{n}}\right)$ are non-differential w.r.t. $\tilde{\mathbf{x}}_{\text{p}}^{k}$, we apply the log-sum-exp (LSE) function to smooth the penalty term following
\begin{equation}
    \max\{a,b\}\approx \rho \ln \left(e^{a/\rho}+e^{b/\rho}\right),    
\end{equation}
where $\rho>0$ denotes the smoothing parameter. As such, $\kappa_{1,k}\left(\tilde{\mathbf{X}}\right)$ and $\kappa_2\left(\tilde{x}_{\text{p}}^{k,{n+1}},\tilde{x}_{\text{p}}^{k,{n}}\right)$ can be approximated as
\begin{equation}
\label{eq:kappa-1-approximation}
    \kappa_{1,k}\left(\tilde{\mathbf{X}}\right)\approx \hat{\kappa}_1\left(\tilde{\mathbf{X}}\right)\triangleq \rho \ln\left(1+e^{\zeta_{1,k}\left(\tilde{\mathbf{X}}\right)/\rho}\right),
\end{equation}
and
\begin{align}
\label{eq:kappa-2-approximation}
    \kappa_2\left(\tilde{x}_{\text{p}}^{k,{n+1}},\tilde{x}_{\text{p}}^{k,{n}}\right)&\approx  \hat{\kappa}_2\left(\tilde{x}_{\text{p}}^{k,{n+1}},\tilde{x}_{\text{p}}^{k,{n}}\right)\nonumber\\
    &\triangleq \rho\ln\left(1+e^{\zeta_2\left(\tilde{x}_{\text{p}}^{k,{n+1}},\tilde{x}_{\text{p}}^{k,{n}}\right)/\rho}\right),
\end{align}
respectively.
Accordingly, we can transform problem (P1-2-2) to the differential problem (P1-2-3) as shown in~\eqref{eq:P1-2-3} at the top of this page. 
\begin{figure*}
    \begin{align}
    (\text{P1-2-3}):&  \max_{\tilde{\mathbf{X}}}  \ q\left(\tilde{\mathbf{X}}\right) \triangleq f\left(\tilde{\mathbf{X}}\right) - \beta\left(\sum\limits_{k=1}^K\hat{\kappa}_1\left(\tilde{\mathbf{X}}\right)+\sum\limits_{k=1}^{K}\sum\limits_{n=1}^{N-1}\hat{\kappa}_2\left(\tilde{x}_{\text{p}}^{k,{n+1}},\tilde{x}_{\text{p}}^{k,{n}}\right)\right).
    \label{eq:P1-2-3}
\end{align}
\hrulefill
\end{figure*}

Note that the equivalence between problem (P1-2-3) and the original problem (P1-2) highly depends on the penalty factor $\beta$ and the LSE smoothing parameter $\rho$. In the following, we first propose the GAA for solving (P1-2-3) with given $\beta$ and $\rho$. Then, the approach for finding the sub-optimal solution of (P1-2) is investigated.
\subsubsection{Solve Problem (P1-2-3) With GAA}
In the $l$-th iteration of the GAA for solving~(P1-2-3), $\tilde{\mathbf{X}}^{(l)}$ is updated by moving along the gradient-ascent direction, i.e.,
\begin{equation}
    \tilde{\mathbf{X}}^{(l+1)} = \tilde{\mathbf{X}}^{(l)} + \tau^{(l)} \left.\frac{\partial q\left(\tilde{\mathbf{X}}\right)}{\partial \tilde{\mathbf{X}}}\right|_{\tilde{\mathbf{X}}=\tilde{\mathbf{X}}^{(l)}},
    \label{eq:gradient-ascent-update}
\end{equation}
{where $\frac{\partial q\left(\tilde{\mathbf{X}}\right)}{\partial \tilde{\mathbf{X}}}\in\mathbb{R}^{K\times N}$, $\tau^{(l)}$ is the step size. The gradient matrix $\frac{\partial q\left(\tilde{\mathbf{X}}\right)}{\partial \tilde{\mathbf{X}}}$ is calculated as follows: }
\begin{equation}
\label{eq:nabla-p}
    \frac{\partial q\left(\tilde{\mathbf{X}}\right)}{\partial \tilde{\mathbf{X}}} = \frac{\partial r\left({\mathbf{X}}\right)}{\partial \mathbf{X}}\circ\frac{L}{2}\left(1-\tanh^2(\tilde{\mathbf{X}})\right),
\end{equation}
where $r\left({\mathbf{X}}\right)$ is the function obtained by substituting $\mathbf{X}=\frac{L}{2}\left(1+\tanh\left(\tilde{\mathbf{X}}\right)\right)$ into $q\left(\tilde{\mathbf{X}}\right)$, and $\circ$ is the Hadamard multiplication. The $(i,j)$-th element of $\frac{\partial r\left(\mathbf{X}\right)}{\partial \mathbf{X}}$, $\forall 1\leq i\leq K, 1\leq j\leq N$ is given in~\eqref{eq:r-partial}, where $\hat{\zeta}_2\left({x}_{\text{p}}^{i,{n+1}},{x}_{\text{p}}^{i,{n}}\right) = \Delta-\left(\mathbf{x}_{\text{p}}^{i,n+1}-\mathbf{x}_{\text{p}}^{i,n}\right)$. For the derivation of $\frac{\partial R_k}{\partial {x}_{\text{p}}^{i,j}}$, the details are given in Appendix~\ref{app:A}. 
\begin{figure*}
\begin{align}
\label{eq:r-partial}
    & \frac{\partial r\left({\mathbf{X}}\right)}{\partial {x}_{\text{p}}^{i,j}} = \sum_{k=1}^{K}\frac{\partial R_k}{\partial {x}_{\text{p}}^{i,j}}-\beta\sum\limits_{k=1}^K\frac{e^{{\zeta_{1,k}}(\mathbf{X})/\rho}}{1+ e^{{\zeta_{1,k}}(\mathbf{X})/\rho}}\frac{\partial {\zeta_{1,k}}(\mathbf{X})}{\partial {x}_{\text{p}}^{i,j}} -\beta \sum_{n=j-1}^{\min\left\{j,N-1\right\}}\frac{e^{\hat{\zeta}_2\left({x}_{\text{p}}^{i,{n+1}},{x}_{\text{p}}^{i,{n}}\right)/\rho}}{1+ e^{\hat{\zeta}_2\left({x}_{\text{p}}^{i,{n+1}},{x}_{\text{p}}^{i,{n}}\right)/\rho}}\frac{\partial \hat{\zeta}_2\left({x}_{\text{p}}^{i,{n+1}},{x}_{\text{p}}^{i,{n}}\right)}{\partial {x}_{\text{p}}^{i,j}}\nonumber\\
    & =\left\{
    \begin{array}{l}
        \sum_{k=1}^{K}\left(1+\beta\frac{e^{{\zeta_{1,k}}(\mathbf{X})/\rho}}{1+ e^{{\zeta_{1,k}}(\mathbf{X})/\rho}}\right)\frac{\partial R_k}{\partial {x}_{\text{p}}^{i,j}} + \beta \frac{e^{\hat{\zeta}_2\left({x}_{\text{p}}^{i,{j}},{x}_{\text{p}}^{i,{j-1}}\right)/\rho}}{1+ e^{\hat{\zeta}_2\left({x}_{\text{p}}^{i,{j}},{x}_{\text{p}}^{i,{j-1}}\right)/\rho}} - \beta\frac{e^{\hat{\zeta}_2\left({x}_{\text{p}}^{i,{j+1}},{x}_{\text{p}}^{i,{j}}\right)/\rho}}{1+ e^{\hat{\zeta}_2\left({x}_{\text{p}}^{i,{j+1}},{x}_{\text{p}}^{i,{j}}\right)/\rho}}, \ \ \ 1\leq j<N,\\
        \sum_{k=1}^{K}\left(1+\beta\frac{e^{{\zeta_{1,k}}(\mathbf{X})/\rho}}{1+ e^{{\zeta_{1,k}}(\mathbf{X})/\rho}}\right)\frac{\partial R_k}{\partial {x}_{\text{p}}^{i,j}} + \beta \frac{e^{\hat{\zeta}_2\left({x}_{\text{p}}^{i,{j}},{x}_{\text{p}}^{i,{j-1}}\right)/\rho}}{1+ e^{\hat{\zeta}_2\left({x}_{\text{p}}^{i,{j}},{x}_{\text{p}}^{i,{j-1}}\right)/\rho}}, \ \ \ j = N.
    \end{array}
    \right.
\end{align}
\hrulefill
\end{figure*}

For the determination of the step size, the backtracking line search is an efficient method~\cite{boyd2004convex}. Specifically, in the $l$-th gradient ascent step, the step size $\tau^{(l)}$ is repeatedly shrunk from the initial value $\bar{\tau}$ with a factor $\omega_{\tau}$, i.e., $\tau^{(l)}\leftarrow \omega_{\tau}\tau^{(l)}$. The searching process terminates when the following Armijo condition is satisfied: 
\begin{equation}
    q\left(\tilde{\mathbf{X}}^{(l+1)}\right) \geq q\left(\tilde{\mathbf{X}}^{(l)}\right) +\delta\tau^{(l)}\left\|\left.\frac{\partial q\left(\tilde{\mathbf{X}}\right)}{\partial\tilde{\mathbf{X}}}\right|_{\tilde{\mathbf{X}}=\tilde{\mathbf{X}}^{(l)}}\right\|_{\text{F}}^2,
    \label{eq:Armijo-condition}
\end{equation}
where $\delta\in\left(0,1\right)$ is the given parameter that controls the increment range of the objective function. The details of the developed algorithm for solving problem (P1-2-3) is shown in \textbf{Algorithm~\ref{alg:GAA}}, where the algorithm terminates when the increment of the objective function in~\eqref{eq:P1-2-3} is less than a given threshold $\epsilon$ or the number of gradient ascent steps reaches the upper bound $l_{\text{max}}$.
\begin{algorithm}[tp]
\caption{Algorithm for Solving Problem (P1-2-1)}
\label{alg:GAA}
\LinesNumbered
\KwIn{$\mathbf{p}$, $\tilde{\mathbf{X}}^{(0)}$, $\beta$, $\rho$, $\bar{\tau}$}
    Set iteration index $l=0$;\\
    \Repeat{$q\left(\tilde{\mathbf{X}}^{(l)}\right) - q\left(\tilde{\mathbf{X}}^{(l-1)}\right)\leq \epsilon$ \textbf{or} $l\geq l_{\text{max}}$}{
    Calculate $\left.{\partial r\left(\mathbf{X}\right)}/{\partial x_{\text{p}}^{i,j}}\right|_{\mathbf{X}=\frac{L}{2}\left(1+\tanh\left(\tilde{\mathbf{X}}^{(l)}\right)\right)}$ in~\eqref{eq:r-partial};\\
    Calculate $\left.{\partial q\left(\tilde{\mathbf{X}}\right)}/{\partial \tilde{\mathbf{X}}}\right|_{\tilde{\mathbf{X}}=\tilde{\mathbf{X}}^{(l)}}$ in~\eqref{eq:nabla-p};\\
    Update $\tau^{(l)} \leftarrow \omega_{\tau}\tau^{(l)}$ until~\eqref{eq:Armijo-condition} is satisfied;\\
    Calculate $\tilde{\mathbf{X}}^{(l+1)}$ in~\eqref{eq:gradient-ascent-update};\\
    $l = l+1$;\\
    }
\textbf{Output} $\mathbf{X} = L/2\left(1+\tanh \left(\tilde{\mathbf{X}}^{(l)}\right)\right)$.
\end{algorithm}
\subsubsection{Find Sub-Optimal Solution for Problem (P1-2)}
Note that the equivalence between the optimality/sub-optimality of problem (P1-2-3) and the original problem (P1-2) highly depends on the values of $\beta$ and $\rho$. It can be verified that, when $\beta\rightarrow +\infty$ and $\rho\rightarrow 0$, the inequality constraints~\eqref{eq:minimum-rate} and~\eqref{eq:antenna-distance} are always guaranteed~\cite{penalty}. However, if the initial value of $\beta$ is set too large, the objective function of~\eqref{eq:P1-2-3} is dominated by the penalty term, and the objective for maximizing the sum rate will be neglected. As such, we first initialize $\beta$ with a small value and gradually $\beta$ to a sufficiently large value. In the mean time, $\rho$ is gradually decreased to improve the approximation accuracy in~\eqref{eq:kappa-1-approximation} and~\eqref{eq:kappa-2-approximation}. As such, the algorithm for solving problem (P1-2) should involve \textbf{Algorithm~\ref{alg:GAA}} iteratively during the update process of $\beta$ and $\rho$, which is summarized in \textbf{Algorithm~\ref{alg:P1-2}}. Specifically, $\beta$ and $\rho$ are updated following $\beta \leftarrow \omega_{\beta}\beta$, $\rho \leftarrow \omega_{\rho}\rho$ in the outer loop, where $\omega_{\beta}>1$ and $0<\omega_{\rho}<1$ are the scaling factors for $\beta$ and $\rho$, respectively. The outer iteration proceeds until constraints~\eqref{eq:minimum-rate} and~\eqref{eq:antenna-distance} are satisfied. 
\begin{algorithm}[tp]
\caption{Algorithm For Solving Problem (P1-2)}
\label{alg:P1-2}
\LinesNumbered
\KwIn{$\mathbf{p}$, $\bar{\tau}$}
Initialize $\tilde{\mathbf{X}}^{(0)}$, $\beta$, and $\rho$;\\
Set iteration index $o=0$;\\
\Repeat{\eqref{eq:minimum-rate} and~\eqref{eq:antenna-distance} are satisfied}{
    Obtain $\tilde{\mathbf{X}}^{(o+1)}$ via \textbf{Algorithm~\ref{alg:GAA}} with given $\beta$ and $\rho$; \\    
    $\beta \leftarrow \omega_{\beta}\beta$, $\rho \leftarrow \omega_{\rho}\rho$; \\
    $o=o+1$; \\
}
\textbf{Output} $\mathbf{X} = L/2\left(1+\tanh \left(\tilde{\mathbf{X}}^{(o)}\right)\right)$.
\end{algorithm}

\subsubsection{Property Analysis of \textbf{Algorithm~\ref{alg:P1-2}}}
Next, we analyze the properties of~\textbf{Algorithm~\ref{alg:P1-2}} in terms of convergence and complexity.
For \textbf{Algorithm~\ref{alg:GAA}}, $q\left(\tilde{\textbf{X}}\right)$ is guaranteed to not decrease after each gradient ascent step, given that $\delta>0$, $\tau^{(l)}>0$ and $\left\|\frac{\partial q\left(\tilde{\mathbf{X}}\right)}{\partial\tilde{\mathbf{X}}}\right\|_{\text{F}}^2\geq 0$. Particularly, if $\left\|\left.\frac{\partial q\left(\tilde{\mathbf{X}}\right)}{\partial\tilde{\mathbf{X}}}\right|_{\tilde{\mathbf{X}}=\tilde{\mathbf{X}}^{(l)}}\right\|_{\text{F}}^2> 0$, we can always find a sufficiently small $\tau^{(l)}$ that satisfies $q\left(\tilde{\mathbf{X}}^{(l+1)}\right) > q\left(\tilde{\mathbf{X}}^{(l)}\right)$. As such, \textbf{Algorithm}~\ref{alg:GAA} is expected to converge to the point where $\frac{\partial q\left(\tilde{\mathbf{X}}\right)}{\partial\tilde{\mathbf{X}}}=0$, which yields the local optimality. Therefore, as $\beta$ approaches infinity, \textbf{Algorithm~\ref{alg:P1-2}} is guaranteed to converge to a stationary point of the original problem (P1-2)~\cite{boyd2004convex}.

The computational complexity of \textbf{Algorithm~\ref{alg:GAA}} mainly relies on the gradient calculation in lines 3-4 and the backtracking line search in line 5. For calculating $\frac{\partial q\left(\tilde{\mathbf{X}}\right)}{\partial \tilde{\mathbf{X}}}$, the complexity is estimated as $\mathcal{O}\left(K^3N\right)$. Denoting the number of iterations for backtracking line search as $I_{\text{ls}}$, the computational complexity for searching the step size is given by $\mathcal{O}\left(I_{\text{ls}}KN\right)$. Therefore, the overall complexity of \textbf{Algorithm~\ref{alg:P1-2}} is estimated as $\mathcal{O}\left(I_{\text{out}}I_{\text{in}}\left(K^3N+I_{\text{ls}}KN\right)\right)$, where $I_{\text{out}}$ and $I_{\text{in}}$ denote the maximum number of outer and inner iterations required for convergence of \textbf{Algorithm~\ref{alg:GAA}} and \textbf{Algorithm~\ref{alg:P1-2}}, respectively.  

\subsection{Pinching Beamforming with Discrete Activation}
Under the practical consideration with limited discrete PAs positions constraint, the pinching beamforming subproblem can be formulated as
\begin{subequations}
\label{eq:discrete-location_optimization_problem}
\begin{equation}
\label{eq:discretelocation_objective-function}
(\text{P1-3}):	\max_{\mathbf{X}}\sum_{k=1}^{K}R_k,
\end{equation}
\begin{equation}
    {\rm{s.t.}} \ \ \eqref{eq:minimum-rate}, 
\end{equation}
\begin{equation}
    \label{eq:discrete-constraint-1}
     {x}_{\text{p}}^{k,n}\in \mathcal{S}_{k}, \forall k, n,
\end{equation}
\begin{equation}
    \label{eq:discrete-constraint-2}
     {x}_{\text{p}}^{k,n}\neq {x}_{\text{p}}^{k,n+1}, \forall k, 1\leq n<N.
\end{equation}
\end{subequations}
Problem (P1-3) is an intractable integer programming problem. To provide a mathematically tractable and low-complexity solution, we propose to utilize the matching theory~\cite{matching1, matching2} to establish the stable matching between the $N$ PAs and $A$ available locations on each waveguide.

\subsubsection{One-Sided One-to-One Matching With Restriction}
\begin{figure}
\centerline{\includegraphics[width=0.9\linewidth]{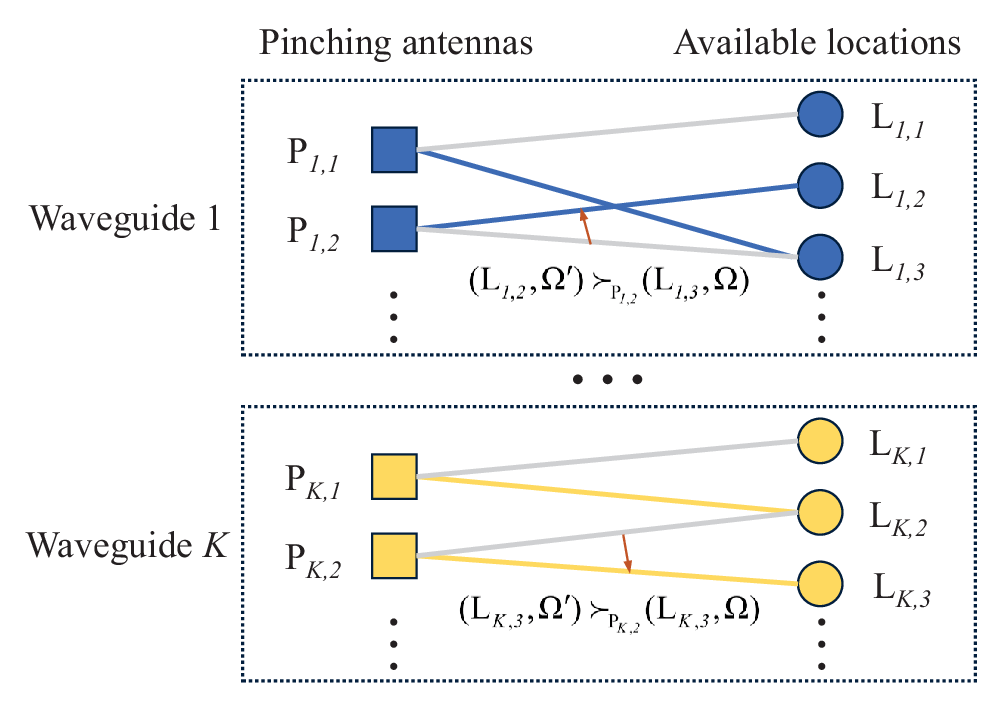}}
\caption{One-to-one matching between PAs and available locations.}
\label{fig:matching-demonstration}
\end{figure}
In the considered matching problem, the set of PAs $\mathcal{P}_{k}=\left\{\text{P}_{k,1}, ..., \text{P}_{k,N}\right\}$ and the set of available locations $\mathcal{L}_{k}=\left\{\text{L}_{k,1}, ..., \text{L}_{k,A}\right\}$, on the waveguide $k$, can be treated as two sets of players that interact with each other to maximize the users' sum rate. Denote the coordinate of the $a$-th location on the $k$-th waveguide, i.e., $\text{L}_{k,a}$, by $\left[\bar{x}_{\text{p}}^{k,a}, y_{\text{p}}^{k}, d\right]$. On each waveguide, a PA should occupy one location, while one location can be assigned to at most one PA. Moreover, PAs have the initiative to choose the matched locations for realizing higher sum rate, while PAs look all the same from the perspective of each location, which means that the formulated problem is an one-sided matching. 

We note that the matching status on waveguide $k$ affects not only the data rate of user $k$, but also the data rate of all other users $k'\neq k$. As such, the matching status of $K$ waveguides are coupled for the maximization of the sum rate. To solve this issue, we propose to formulate the discrete PAs positions optimization problem over all $K$ waveguides as a one-sided one-to-one matching problem, where the two sets of players are denoted by $\mathcal{P}=\left\{\mathcal{P}_1, ..., \mathcal{P}_K\right\}$ and $\mathcal{L}=\left\{\mathcal{L}_1, ..., \mathcal{L}_K\right\}$. Note that, unlike the conventional matching model, where one set of the players can be matched to any players in the other set, our proposed matching model has the \textit{restriction} on the selectable matching partners. Specifically, PAs in $\mathcal{P}_k$ can only be matched to available locations in $\mathcal{L}_k$, as shown in Fig.~\ref{fig:matching-demonstration}.
Accordingly, we give the definition of the proposed one-sided one-to-one matching with \textit{restriction} as follows:   
\begin{definition}
    In the one-sided one-to-one matching model with restriction, a matching $\Omega$ is a mapping function from $\mathcal{P}$ to $\mathcal{L}$ that satisfies
    
    1) $\Omega\left(\text{P}_{k,n}\right)\in \mathcal{L}_k, \forall k,n$; 
    
    2) 
    $\left|\Omega\left(\text{P}_{k,n}\right)\right|=1, \forall k,n$; 

    3) $\Omega\left(\text{P}_{k,n}\right)=\text{L}_{k,a}\Rightarrow \Omega\left(\text{P}_{k,n'}\right)\neq \text{L}_{k,a}, \forall k, n\neq n'$.
\end{definition}

In the proposed matching game, another critical concept is the \textit{preference} of each PA on the available positions, which is determined by the achievable utility. Since our objective in problem (P1-3) is to maximize the sum rate, we define the utility function as $U\left(\Omega\right) = \sum_{k=1}^KR_k$, where $U\left(\Omega\right)$ is the utility under the matching state $\Omega$. The preference list of each PA is formed in descending order w.r.t. the utility achieved by the matched positions. 

\begin{remark}
    The proposed matching game has external effect~\cite{matching3}. That is, from the perspective of $\text{P}_{k,n}, \forall k, n$, the utility not only depends on the identity of its matched partner, but also relies on where other PAs $\text{P}_{k',n'}, \forall k'\neq k, n'\neq n$ are matched.
\end{remark}
With the external effect, each PA has a dynamic preference list relying on the matching state rather than a fixed one. Accordingly, we say the PA $\text{P}_{k,n}$ \textit{prefers}, denoted by ``$\succ$", location $\text{L}_{k,a}$ rather than its currently matched position under the state $\Omega$, when the following condition is satisfied:
\begin{equation}
    \left(\text{L}_{k,a}, \Omega'\right) \succ_{\text{P}_{k,n}} \left(\Omega\left(\text{P}_{k,n}\right), \Omega\right)\Leftrightarrow U\left(\Omega'\right)>U\left(\Omega\right),
\end{equation}
where $\left(\text{L}_{k,a}, \Omega'\right)$ represents that $\Omega'\left(\text{P}_{k,n}\right)=\text{L}_{k,a}$. $\Omega'$ is the updated matching state obtained by pairing $\text{P}_{k,n}$ with $\text{L}_{k,a}$, which is given by
\begin{equation}
    \Omega'=\Omega\backslash \left\{\left(\text{P}_{k,n}, \Omega\left(\text{P}_{k,n}\right)\right)\right\}\cup \left\{\left(\text{P}_{k,n}, \text{L}_{k,a}\right)\right\}.
\end{equation}
With the preference of PAs over available locations, the definition of a \textit{stable} maching is given below.
\begin{definition}
    Based on the proof provided in~\cite{matching3}, a matching state $\Omega$ is defined as a \textit{stable} matching when there does not exist a matching $\Omega'$ that satisfies 
    
    1) $\left(\Omega'\left(\text{P}_{k,n}\right), \Omega'\right) \succeq_{\text{P}_{k,n}} \left(\Omega\left(\text{P}_{k,n}\right), \Omega\right), \forall k, n$; 
    
    2)  $\left(\Omega'\left(\text{P}_{k,n}\right), \Omega'\right) \succ_{\text{P}_{k,n}} \left(\Omega\left(\text{P}_{k,n}\right), \Omega\right), \exists k, n$.
\end{definition}
\subsubsection{Proposed PAs positions Optimization Algorithm Based on One-to-One Matching}
To solve the discrete PAs positions optimization problem (P1-3), we propose a matching algorithm as detailed in~\textbf{Algorithm~\ref{alg:matching}}, for finding the stable matching between PAs and available locations. Specifically, we first initialize the matching state, where PAs are randomly assigned to different locations on the corresponding waveguide, as shown in line 1. Subsequently, in line 4-9, each PA searches for all available locations on the corresponding waveguide, to check whether there is an unoccupied location given higher preference compared to the currently matched partner. This searching process continues until there is no PA can find any preferred location, as shown in line 10. 
\begin{algorithm}[tp]
	\caption{Algorithm For Solving Problem (P1-3)}
    \label{alg:matching}
	\LinesNumbered
	\KwIn{$\mathbf{p}$}
	Initialize matching state $\Omega^{(0)}$: randomly match each PA $\text{P}_{k,n}$ with a location in $\mathcal{L}_k$ with the condition on satisfying constraint~\eqref{eq:minimum-rate}, where each location can only be matched to one PA\;
    $\Omega = \Omega^{(0)}$\;
    \Repeat{$\Omega == \bar{\Omega}$}{
        $\bar{\Omega} = \Omega$\;
        \For{$k = 1:K$}{
            \For{$n = 1:N$}{
                \For{$a = 1:A$}{
                    \If{$\Omega\left(\text{P}_{k,n'}\right) \neq \text{L}_{k,a}, \forall n' \text{ and } \left(\text{L}_{k,a}, \Omega'\right) \succ_{\text{P}_{k,n}} \left(\Omega\left(\text{P}_{k,n}\right), \Omega\right)$ \text{ and } \eqref{eq:minimum-rate} is satisfied}{
                        $\Omega = \Omega'$;
                    }
                }
            }
        }
    }
    \KwOut{Final matching $\Omega^*$}
\end{algorithm}

\subsubsection{Property Analysis of~\textbf{Algorithm~\ref{alg:matching}}}
To evaluate the performance of~\textbf{Algorithm~\ref{alg:matching}}, we analyze its properties in terms of complexity, convergence, stability and optimality in the following. 
\begin{itemize}
    \item \textit{Complexity}: As shown in~\textbf{Algorithm~\ref{alg:matching}}, the complexity of the proposed matching algorithm mainly relies on the number of iterations for searching for the ``PA-location" pairs that can further improve the sum rate in line 3-10. As such, denoting the maximum number of the outermost cycle by $I_{\text{mat}}$, the complexity of~\textbf{Algorithm~\ref{alg:matching}} is estimated as $\mathcal{O}\left(I_{\text{mat}}KNA\right)$, which increase linearly with $K$, $N$, and $A$, respectively.
\item \textit{Convergence and Optimality}: After obtaining the initial matching, \textbf{Algorithm~\ref{alg:matching}} attempts to re-locate PAs for improving the sum rate. Denote the update process of matching states as ``$\Omega^{(0)}\rightarrow \Omega^{(1)}\rightarrow \dots \rightarrow \Omega^*$". Then we can easily prove that 
\begin{equation}
\label{eq:utility-change}
    U\left(\Omega^{(0)}\right) < U\left(\Omega^{(1)}\right) < \dots < U\left(\Omega^{*}\right).
\end{equation}
One can observe from~\eqref{eq:utility-change} that the users' sum rate keeps increasing with the update of the matching state. Since the sum rate is upper bounded with the limited power, \textbf{Algorithm~\ref{alg:matching}} is guaranteed to coverge within limited number of iterations. Moreover, since the final matching $\Omega^*$ represents a state where no PA can be re-located to a different location for further improving the utility, which yields a local optimum for problem (P1-3).
\item \textit{Stability}: 
Assume that the final matching $\Omega^*$ of~\textbf{Algorithm~\ref{alg:matching}} is not stable. In other words, there exists a matching $\Omega'\neq \Omega^*$ satisfying that  $\forall k, n, \left(\Omega'\left(\text{P}_{k,n}\right), \Omega'\right) \succeq_{\text{P}_{k,n}} \left(\Omega^*\left(\text{P}_{k,n}\right), \Omega^*\right)$ and $\exists k, n, \left(\Omega'\left(\text{P}_{k,n}\right), \Omega'\right) \succ_{\text{P}_{k,n}} \left(\Omega\left(\text{P}_{k,n}\right), \Omega\right)$. However, the termination condition of the outmost cycle in~\textbf{Algorithm~\ref{alg:matching}} is that all PAs can not find any available location that achieves utility increment compared to the currently matched location. This means that $\Omega^*$ is not the final matching, which causes conflict. Therefore, we can say that the output of~\textbf{Algorithm~\ref{alg:matching}} is a stable matching. 
\end{itemize}
\subsection{Overall AO Algorithm Design}
Based on the results obtained in the previous three subsections, we now finalize the overall algorithm for solving the original problem (P1) based on the AO method, which is presented in details in ~\textbf{Algorithm~\ref{alg:AO}}. Specifically, the PAs positions $\mathbf{X}$ and the power allocation $\mathbf{p}$ are alternatively optimized, by solving problem (P1-1) and (P1-2) (or (P1-3)) correspondingly, while keeping the other variables fixed. The iterations terminate when the increment of the sum rate is below the predefined threshold $\epsilon$. Since the sum rate is non-decreasing for both the optimization of $\mathbf{X}$ and $\mathbf{p}$, the convergence of \textbf{Algorithm~\ref{alg:AO}} is guaranteed due to the upper-bounded sum rate.


\begin{algorithm}[tp]
	\caption{Algorithm For Solving Problem (P1)}
        \label{alg:AO}
	\LinesNumbered
	\KwIn{$\bar{\tau}$}
	Initialize $\mathbf{p}^{(0)}$;\\
    Set iteration index $r=0$;\\
    \Repeat{Increment of the sum rate is below $\epsilon$}{
    Solve problem (P1-2) (or (P1-3)) for given $\mathbf{p}^{(r)}$ with~\textbf{Algorithm~\ref{alg:P1-2}} (or \textbf{Algorithm~\ref{alg:matching}}), and denote the solution as $\mathbf{X}^{(r+1)}$;\\
    Solve problem (P1-1) for given $\mathbf{X}^{(r+1)}$ with~\textbf{Algorithm~\ref{alg:SCA}}, and denote the solution as $\mathbf{p}^{(r+1)}$;\\
    $r=r+1$;
    }
    \KwOut{$\mathbf{X}, \mathbf{p}$}
\end{algorithm}

\section{Simulation Results}
In this section, numerical results are provided to validate the feasibility and effectiveness of the proposed PASS-enabled multi-user communications design with the WDMA.

\subsection{Simulation Setup}
In simulations, we focus on the case where two waveguides are employed to serve two users, i.e., $K=2$. Referring to the coordinate system in Fig.~\ref{fig:system_model}, we assume that all waveguides are with the same height of $d=3$~meters, each covering a rectangular area with size $W\times L$. The waveguides are set with the distance of $W$ along the $y$-axis, and the coordinates of the feed points on the two waveguides are $\left(0,-\frac{W}{2}, 3\right)$~meters and $\left(0,\frac{W}{2}, 3\right)$~meters, respectively. The first user is randomly distributed within the upper rectangle, while the second user is randomly distributed within the lower rectangle. Unless otherwise specified, the adopted system parameters are presented in Table~\ref{table:parameters}. Each point in Figs.~7 - 11 is obtained by averaging over $500$ channel realizations generated by randomly distributing the two users within the coverage areas of their serving waveguides, while the simulation results in Fig.~6 are obtained over one random channel realization for illustrating the convergence performance of the proposed algorithms.
\begin{table}[h] 
\caption{Simulation parameters}
\centering
\label{table:parameters}
\begin{tabular}{|p{1cm}|p{4.5cm}|p{1.5cm}|}
\hline
Parameter & Description & Value\\
\hline
$f$ & Carrier frequency & $28$~GHz \\ \hline
{$\Delta$} & Minimum PAs distance  & $\frac{\lambda}{2}$\\ \hline
$\bar{R}_k$ & Minimum rate requirement of each user & $2$~bps/Hz\\ \hline
$n_{\text{eff}}$ & Effective refractive index of a dielectric waveguide & $1.4$\\ \hline
$\sigma_k^2$ & User noise power  & $-90$~dBm\\ \hline
$P_{\text{max}}$ & BS maximum transmit power & $20$~dBm\\ \hline
$W, L$ & Side length of each waveguide's coverage area & $6$~m, $10$~m\\ \hline
$\overline{\tau}$ & Initial gradient-ascent step size in Algorithm 2 & $10$ \\ \hline
$\omega_{\tau}$ & Step size shrinking parameter in Algorithm 2 & $0.5$ \\ \hline
$\epsilon$ & Convergence threshold in Algorithm 1,2,5 & $10^{-6}$ \\ \hline
$\beta$ & Initial penalty factor in Algorithm 3 & $10^{-4}$ \\ \hline
${\rho}$ & Initial smoothing parameter in Algorithm 3 & $1$ \\ \hline
$\omega_{\beta}$ & Scaling factor in Algorithm 3 & $10$ \\ \hline
$\omega_{\rho}$ & Scaling factor in Algorithm 3 & $0.1$ \\ \hline
\end{tabular}
\end{table}
\begin{figure*}
	\begin{subfigure}
	    {0.32\linewidth}
		\centerline{\includegraphics[width=\textwidth]{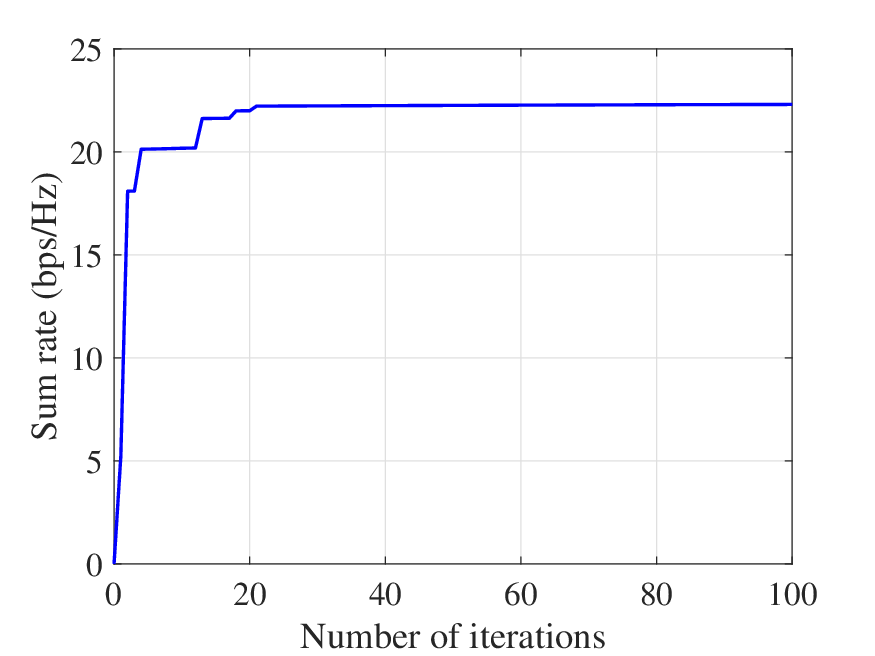}}
		\caption{Convergence performance of Algorithm~\ref{alg:P1-2} with $N=2$.}
        \label{fig:algorithm2-convergence}
	\end{subfigure}
	\begin{subfigure}{0.32\linewidth}
		\centerline{\includegraphics[width=\textwidth]{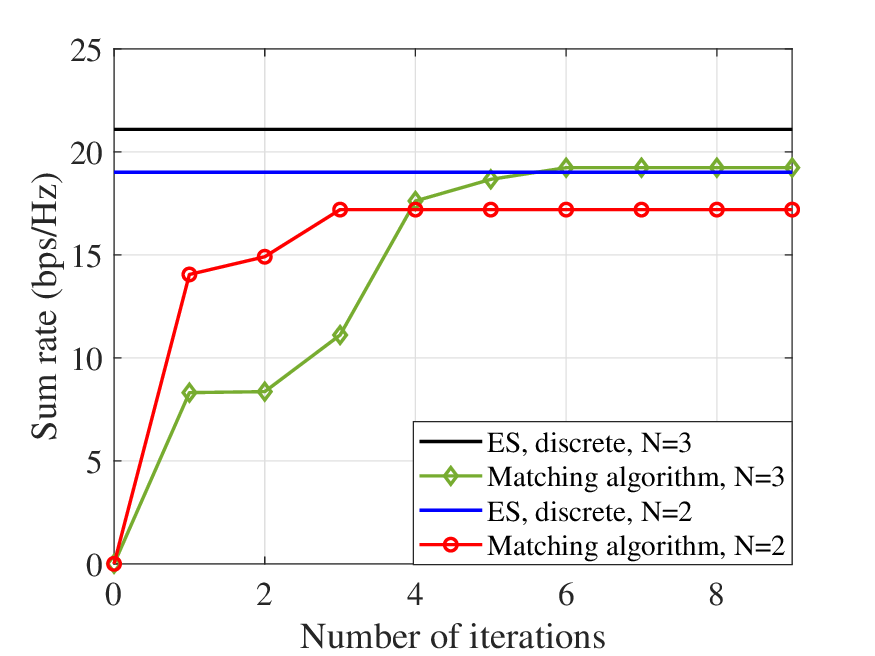}}
		\caption{Convergence performance of Algorithm~\ref{alg:matching} with $A=20$.}
        \label{fig:algorithm5-convergence}
	\end{subfigure}
	\begin{subfigure}{0.32\linewidth}
		\centerline{\includegraphics[width=\textwidth]{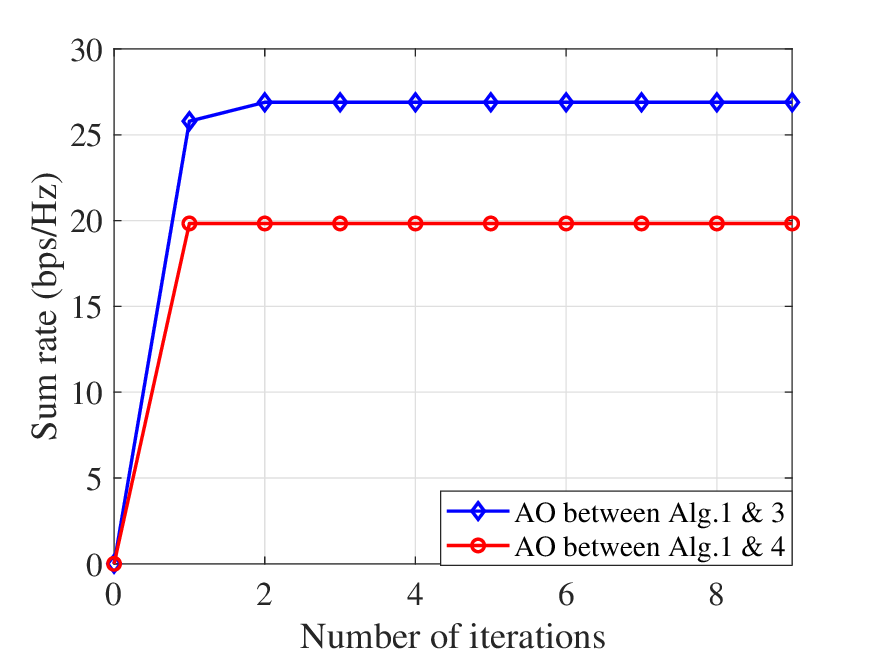}}
		\caption{Convergence performance of Algorithm~\ref{alg:AO} with $N=4$.}
	\end{subfigure}
    \caption{Convergence behavior of the proposed algorithms.}
	\label{fig:convergence}
    \vspace{-0.2cm}
\end{figure*}
\subsection{Baseline Schemes}
To verify the effectiveness of the proposed multi-user communications scheme in PASS, we compare with the following baseline schemes. 
\begin{itemize}
    \item \textbf{Conventional antenna systems}: In this case, two fixed-position BS antennas are deployed right above the centroid of the considered area, where the coordinates of two antennas are $\left(\frac{L-\Delta}{2}, 0, 3\right)$~meters and $\left(\frac{L+\Delta}{2}, 0, 3\right)$~meters, respectively. The fully-digital beamforming is assumed to be enabled at the BS. Note that, since multiple PAs can be activated on a waveguide at no extra cost, we assume the number of antennas in conventional antenna systems is same as the number of waveguides in the considered PASS for fairly comparing the performance under the same installation cost. 

    \item \textbf{Maximum ratio transmission based continuous antenna activation (``MRT, continuous")}: The basic idea of this baseline scheme is similar to that of the MRT beamforming in the MIMO communications system, where the antennas on each waveguide are activated aiming for maximizing the desired signal strength at the corresponding user. To find the desirable antennas locations, the two-stage algorithm proposed in~\cite{yanqing} is adopted. Specifically, in the first stage, the locations are optimized in a large scale for path loss minimization of the desired signals. In the second stage, the location are further refined in the wavelength scale, such that desired signals received at the users can be constructively combined. 
    For more details, we refer the reader to~\cite[Section III]{yanqing}. 
    \item \textbf{GAA discretization based discrete antenna activation (``GAA, discrete")}: In this case, the continuous antennas locations obtained by the GAA algorithm proposed in Section III-A are discretized to the closest feasible location for activating the antennas. If the approximated discrete locations of two PAs are the same, the antenna with a larger difference between the derived and the approximated locations values will be activated at the next nearest location that is unoccupied by any other antenna.
    \item \textbf{Exhaustive search based discrete antenna activation (``ES, discrete")}: In this case, the exhaustive search is performed through searching all possible discrete activation locations for antennas, so as to obtain the optimal solution that achieves the highest sum rate. This baseline scheme can provide the performance upper bound for the discrete activation case. 
\end{itemize}

\subsection{Convergence of The Proposed Algorithms}
In Fig.~\ref{fig:convergence}, we first provide the convergence performance of \textbf{Algorithm~\ref{alg:P1-2}}, \textbf{Algorithm~\ref{alg:matching}}, and \textbf{Algorithm~\ref{alg:AO}}.  As can be observed in Fig.~6(a), for the case of continuous activation with $N=2$, the sum rate achieved by the proposed GAA keeps increasing in the first few iterations and converges to a steady point after around $30$ iterations. For the case of discrete activation with $A=20$, Fig.~6(b) demonstrates that the proposed matching algorithm can converge within around $3$ and $6$ outermost iterations, i.e., $I_{\text{mat}}$, for the case of $N=2$ and $N=3$, respectively. It is observed that the sum rate increases with larger $N$. This is expected, as a larger number of PAs bring in higher power gain. However, the convergence speed gets slower when the number of antennas increases, given that the matching process gets more complicated with more involved players. It is also shown that \textbf{Algorithm~\ref{alg:matching}} achieves nearly $91\%$ of the performance upper bound obtained via ``ES, discrete". Recall the computational complexity analysis of \textbf{Algorithm~\ref{alg:matching}} in Section III-C. We note that, \textbf{Algorithm~\ref{alg:matching}} can achieve near-optimal solution with low complexity.  

Furthermore, we depict the convergence performance of the AO algorithm in Fig.~6(c), where both the alternating optimization process between \textbf{Algorithm~\ref{alg:SCA}} and \textbf{Algorithm~\ref{alg:P1-2}} as well as that between \textbf{Algorithm~\ref{alg:SCA}} and \textbf{Algorithm~\ref{alg:matching}} are demonstrated. We set $N=4$ for all considered schemes and $A=20$ for discrete activation schemes. It can be seen that the alternation between the power allocation and the pinching beamforming can converge within very limited number of iterations, which implies the possible application of the proposed AO algorithm in practice. 

\subsection{Sum Rate Versus BS Maximum Transmit Power}
\begin{figure}
    \centering
    \centerline{\includegraphics[scale=0.50]{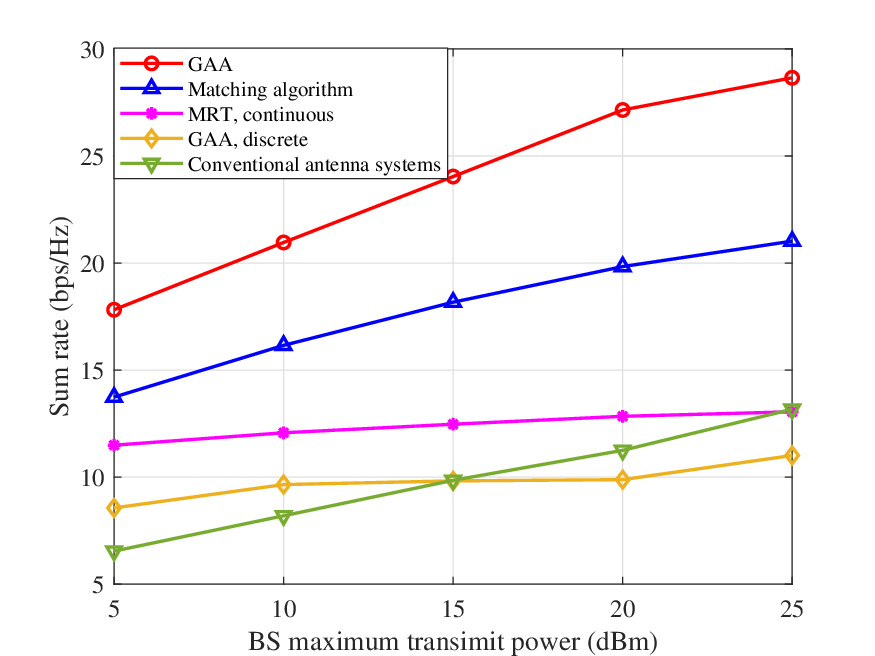}}
    \caption{Sum rate versus BS maximum transmit power $P_{\text{max}}$ with $N=4$ and $A=20$ for discrete activation schemes.}
    \label{fig:sum-rate-power}
\end{figure}

In Fig.~\ref{fig:sum-rate-power}, we compare the achievable sum rate of different schemes versus the BS maximum transmit power $P_{\text{max}}$. We set $N=4$ for all considered schemes and $A=20$ for discrete activation schemes. As can be seen from the figure, the considered PASS significantly outperforms conventional antenna systems in both the continuous and discrete activation cases. The reasons behind this can be explained as follows. On the one hand, PAs on each waveguide can be flexibly activated close to the target user to build a stronger LoS link, while the conventional antennas can only be deployed at the BS side and thus leads to high path loss. On the other hand, since multiple PAs can be activated on a single waveguide at no extra cost, the higher number of antennas in PASS brings in more degree-of-freedoms (DoFs) to pinching beamforming. These insights underscore the superiority of the WDMA for underpinning multi-user communications in a new paradigm.

Regarding the effectiveness of the proposed GAA for continuous activation, {one can observe from Fig.~\ref{fig:sum-rate-power} that the GAA significantly outperforms the ``MRT, continuous".} This can be explained as follows. With the objective of sum rate maximization, the results obtained by GAA is a trade-off between maximizing the intended users' signal strength and minimizing the inter-user interference. For the ``MRT, continuous", the intended users' signal strength is maximized at the price of uncontrolled interference, which results in the performance that is even inferior to conventional antenna systems. It is also observed that the proposed matching algorithm significantly outperforms ``GAA, discrete". This implies the importance of the specialized optimization scheme design tailored for the discrete activation case. 

\subsection{Effectiveness of the Pinching Beamforming}
\begin{figure}
	\centering
	\begin{subfigure}{1.0\linewidth}
		\centering
		\includegraphics[width=0.75\linewidth]{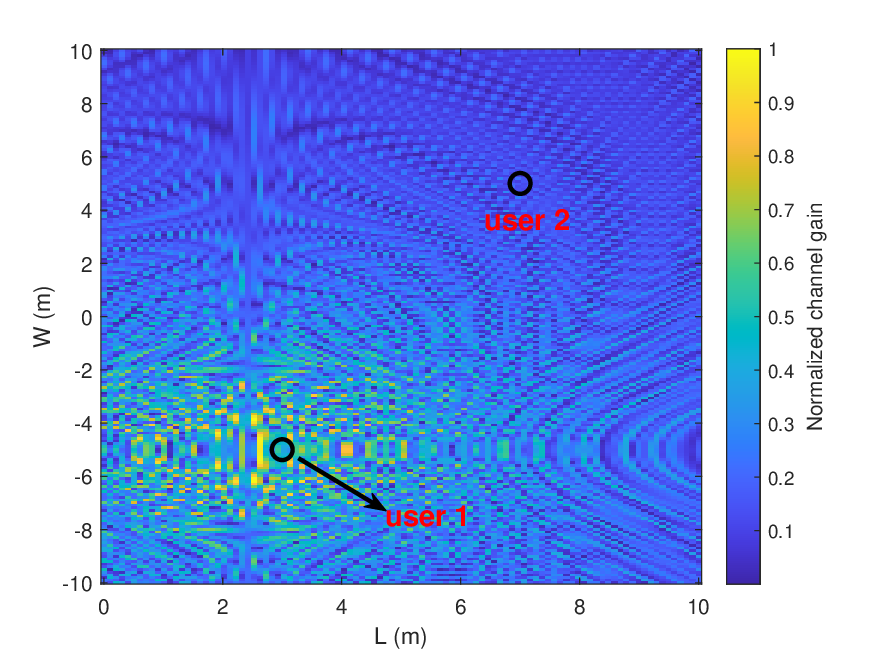}
		\caption{Waveguide 1.}
		\label{fig:channel-gain-waveguide-1}
	\end{subfigure}
	\centering
	\begin{subfigure}{1.0\linewidth}
		\centering
		\includegraphics[width=0.75\linewidth]{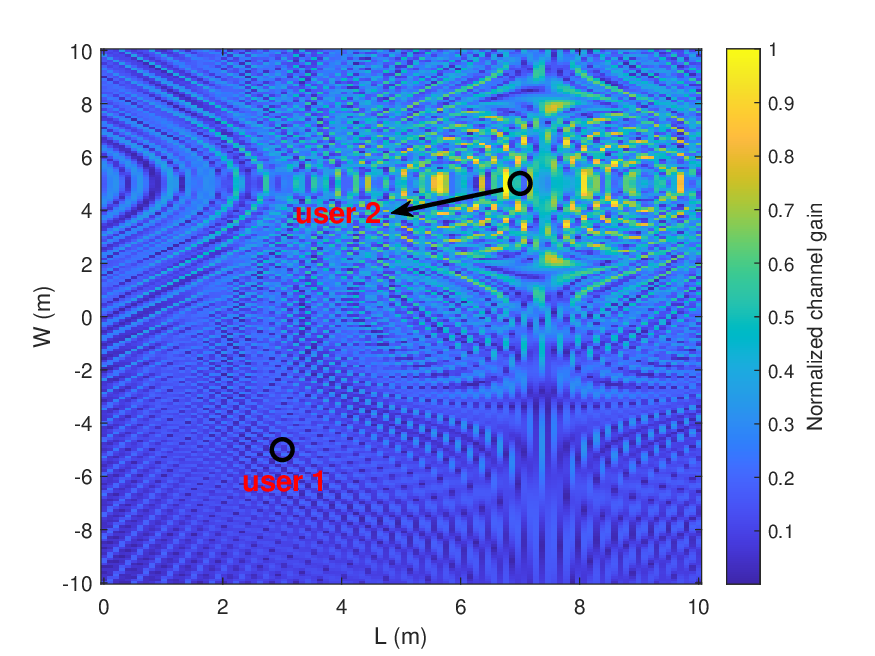}
		\caption{Waveguide 2.}
		\label{fig:channel-gain-waveguide-2}
	\end{subfigure}
    \caption{Illustration of the channel gain from each waveguide over the serving area with user 1 and user 2 located at $(3,-5,0)$~meters and $(7,5,0)$~meters, respectively.}
	\label{fig:channel-gain}
\end{figure}

Fig.~\ref{fig:channel-gain} illustrates the normalized channel gain from each waveguide over the coverage area, where user 1 and 2 are located at $(3,-5,0)$~meters and $(7,5,0)$~meters, respectively. The normalized channel gain is calculated with the PAs positions obtained by the proposed GAA, where we set $N=4$. 
As shown in Fig.~\ref{fig:channel-gain}, for each waveguide, high channel gain can be achieved for the intended user, while much lower channel gain is observed at the interfering user. This implies that the nearly orthogonal transmission can be realized via the pinching beamforming, which verifies the effectiveness. 

\subsection{Sum Rate Versus Number of PAs on Each Waveguide}
\begin{figure}
    \centering
    \centerline{\includegraphics[scale=0.50]{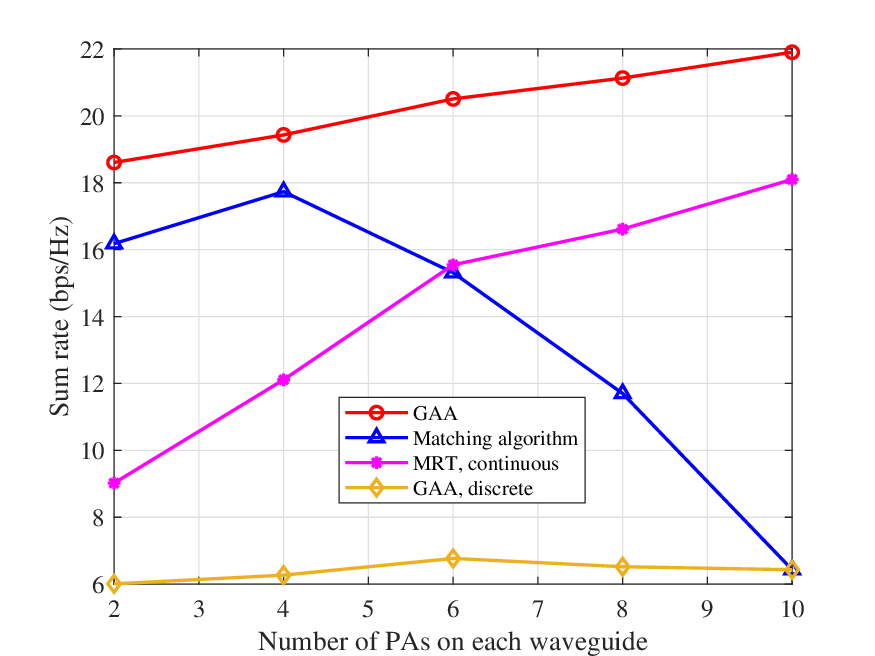}}
    \caption{Sum rate versus number of PAs on each waveguide $N$, where $A=10$.}
    \label{fig:sum-rate-N}
\end{figure}

Fig.~\ref{fig:sum-rate-N} depicts the sum rate versus the number of PAs on each waveguide $N$. We set $A=10$ for discrete activation schemes. It is observed that the sum rate for both GAA and ``MRT, continuous" increases with the increment of $N$. This is expected, as larger $N$ brings in higher spatial DoFs for the pinching beamforming, thereby improving the multi-user sum rate. Furthermore, it is interesting to find that, for both matching algorithm and ``GAA, discrete", the sum rate increases firstly and decreases afterwards with the increment of $N$. This is due to the fact, since the number of candidate positions is limited, the optimal sum rate is confined by the set of possible ``PA-position" pairs. As such, the performance gain brought by the increasing number of PAs is restricted. When $N$ reaches to a certain value, the increment of the number of PAs can result in the decrement of the sum rate for forcing to place the PAs on the inappropriate positions. For example, when $N=10$, the performance of the proposed matching algorithm and ``GAA, discrete" are the same, as the PA activation strategies obtained by these two schemes are both to place PAs on all the available positions. 
On the contrary, since the PAs positions can be adjusted in a fine manner in the continuous case, the higher flexibility brings in more pronounced performance gain compared to the discrete case for larger $N$. The obtained results highlight the importance for determining the proper number of activated PAs in the discrete case, which can be left for future work. 
\subsection{Sum Rate Versus Region Size}
\begin{figure}
    \centering
    \centerline{\includegraphics[scale=0.50]{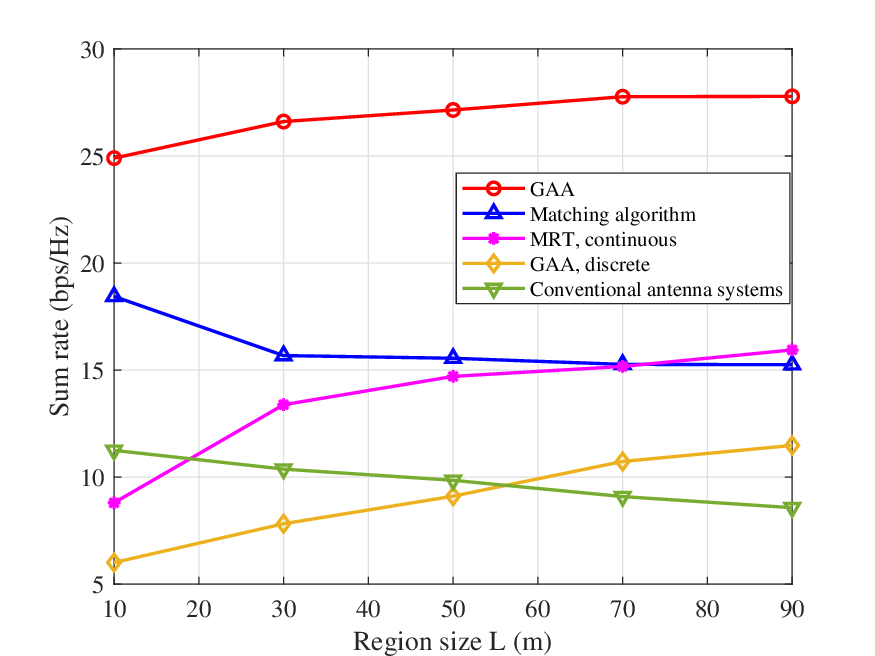}}
    \caption{Sum rate versus region size $L$, where $N=4$.}
    \label{fig:sum-rate-L}
\end{figure}


In Fig.~\ref{fig:sum-rate-L}, we investigate the achievable sum rate versus the region size $L$. We set $N=4$ for all considered schemes and $A=20$ for discrete activation schemes. It is interesting to find that the sum rate achieved by the GAA increases by enlarging $L$, while that of conventional antenna systems decreases with the increment of $L$. This is because, on the one hand, since PAs can be flexibly activated near to the intended users, larger $L$ leads to higher sum rate in PASS, thanks to the reduced interference among the randomly distributed users. On the other hand, the increased users' distance to conventional antennas leads to higher path loss on average in conventional antenna systems. This performance comparison confirms the superiority of PASS especially in extended service areas. 

Moreover, we also notice that sum rate achieved by the proposed matching algorithm decreases with larger $L$. Since the number of candidate positions is fixed, larger $L$ leads to lower discretization accuracy and inferior performance.  In addition, the sum rate achieved by ``MRT, continuous" increases with the increment of $L$, which can be explained as follows. The objective of ``MRT, continuous" is to maximize the intended user signal strength while ignoring the impact of the pinching beamforming on the interference. Since the multi-user interference degrades with the increment of $L$, higher sum rate can be obtained with ``MRT, continuous".


\subsection{Impact of Discrete PA Activation}
\begin{figure}
    \centering
    \centerline{\includegraphics[scale=0.50]{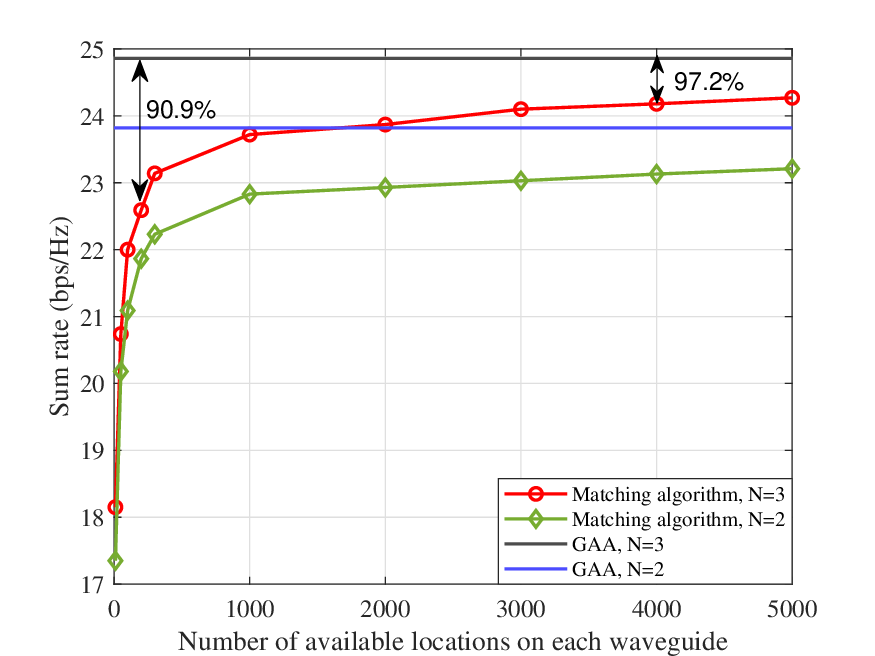}}
    \caption{Sum rate versus number of available positions $A$ in the discrete case.}
    \label{fig:sum-rate-A}
\end{figure}
In Fig.~\ref{fig:sum-rate-A}, we present the impact of the number of candidate positions on the sum rate performance of discrete PA activation schemes. One can first observe that the sum rate increases with larger $A$. This is expected, as PAs can be activated more precisely when there exists more available positions. It is also seen that, the performance gap between the continuous and discrete cases is alleviated with the increment of $A$. For example, the sum rate of the proposed matching algorithm can achieve around $90\%$ and $97\%$ to that of the GAA for $A=200$ and $A=5000$, respectively. Since the computational complexity is enlarged with higher discretization accuracy, the tradeoff between the performance and the complexity should be taken into consideration for the practical implementation.

\section{Conclusions}
We proposed a simple multi-user communications design for employing PASS with the novel concept of WDMA. For both ideal continuous and practical discrete PA activation cases, the joint power allocation and pinching beamforming problem was formulated for maximization of the users' sum rate. By employing the AO algorithm, the joint problem was effectively decomposed into two subproblems. The SCA method was first developed for solving the power allocation subproblem. Subsequently, the penalty-based GAA and the matching algorithm were invoked for solving the pinching beamforming subproblems in the continuous and discrete activation cases, respectively. 
Simulation results revealed that, with the exploitation of the WDMA, PASS significantly improved the system sum rate compared to conventional antenna systems. Furthermore, it was also shown that, the performance gap between the discrete and continuous activation cases could be alleviated by increasing the number of available positions. 


\begin{appendices} 
\section{Derivation of ${\partial R_k}/{\partial x_{\text{p}}^{i,j}}$}
\label{app:A}
For ease of brevity, define $\upsilon_{k',k}=\left|\sum\limits_{n=1}^N\frac{\eta e^{-j\phi_{k',k}^n}}{\left\|\boldsymbol{\psi}_{\text{u}}^{k}-\boldsymbol{\psi}_{\text{p}}^{k',n}\right\|}\right|^2$. In the following, we disucss two different cases for the derivation of ${\partial R_k}/{\partial x_{\text{p}}^{i,j}}$. 
\subsubsection{$i = k$}
If $i=k$, we obtain ${\partial R_k}/{\partial x_{\text{p}}^{i,j}}$ as
\begin{equation}
    \frac{\partial R_k}{\partial x_{\text{p}}^{i,j}}=\frac{p_i}{\ln2\left(\sum\limits_{k'\neq i}^Kp_{k'}\upsilon_{k',k}+N\sigma_{k}^2+p_i\upsilon_{i,k}\right)}\times\frac{\partial \upsilon_{i,k}}{\partial x_{\text{p}}^{i,j}}.
\end{equation}

Let $\upsilon_{i,k}=\left|\frac{\eta e^{-j\phi_{i,k}^j}}{\left\|\boldsymbol{\psi}_{\text{u}}^k-\boldsymbol{\psi}_{\text{p}}^{i,j}\right\|}+\xi_{i,k}^j\right|^2$, where $\xi_{i,k}^j \triangleq\sum\limits_{n\neq j}^N\frac{\eta e^{-j\phi_{i,k}^n}}{\left\|\boldsymbol{\psi}_{\text{u}}^k-\boldsymbol{\psi}_{\text{p}}^{i,n}\right\|}=\left|\xi_{i,k}^j\right|e^{j\angle\xi_{i,k}^j}$. We can then rewrite $\upsilon_{i,k}$ as follows:
\begin{align}
    &\upsilon_{i,k}=\left|\frac{\eta e^{-j\left(\frac{2\pi}{\lambda}c_{i,k}^j+\frac{2\pi}{\lambda_g}x_{\text{p}}^{i,j}\right)}}{c_{i,k}^j}+\xi_{i,k}^j\right|^2\nonumber\\
    & = \left(\frac{\eta\cos{\left(\frac{2\pi}{\lambda}c_{i,k}^j+\frac{2\pi}{\lambda_g}x_{\text{p}}^{i,j}\right)}}{c_{i,k}^j}+\left|\xi_{i,k}^j\right|\cos{\angle \xi_{i,k}^j}\right)^2\nonumber\\
    & +\left(-\frac{\eta\sin{\left(\frac{2\pi}{\lambda}c_{i,k}^j+\frac{2\pi}{\lambda_g}x_{\text{p}}^{i,j}\right)}}{c_{i,k}^j}+\left|\xi_{i,k}^j\right|\sin{\angle \xi_{i,k}^j}\right)^2,
\end{align}
where $c_{i,k}^j=\left\|\boldsymbol{\psi}_{\text{u}}^k-\boldsymbol{\psi}_{\text{p}}^{i,j}\right\|=\sqrt{\left(x_{\text{p}}^{i,j}-x_{\text{u}}^{k}\right)^2+\left(y_{\text{p}}^{i}-y_{\text{u}}^{k}\right)^2+d^2}$. Then, we obtain ${\partial \upsilon_{i,k}}/{\partial x_{\text{p}}^{i,j}}$ as given in~\eqref{eq:partial-upsilon}.
\begin{figure*}
    \begin{align}
    \label{eq:partial-upsilon}
        \frac{\partial \upsilon_{i,k}}{\partial x_{\text{p}}^{i,j}} =& -\frac{2\eta^2\left(x_{\text{p}}^{i,j}-x_{\text{u}}^k\right)}{\left(c_{i,k}^j\right)^4}-\frac{2\left|\xi_{i,k}^j\right|\eta\cos{\left(\phi_{i,k}^j+\angle \xi_{i,k}^j\right)}\left(x_{\text{p}}^{i,j}-x_{\text{u}}^k\right)}{\left(c_{i,k}^j\right)^3}\nonumber\\
        & -\frac{4\pi\left|\xi_{i,k}^j\right|\eta\sin{\left(\phi_{i,k}^j+\angle \xi_{i,k}^j\right)}\left(x_{\text{p}}^{i,j}-x_{\text{u}}^k\right)}{\lambda\left(c_{i,k}^j\right)^2}-\frac{4\pi\left|\xi_{i,k}^j\right|\eta\sin{\left(\phi_{i,k}^j+\angle \xi_{i,k}^j\right)}}{\lambda_gc_{i,k}^j}.
    \end{align}
\end{figure*}

\subsubsection{$i\neq k$} If $i\neq k$, we obtain ${\partial R_k}/{\partial x_{\text{p}}^{i,j}}$ as given in~\eqref{eq:partial-R} following the similar derivations for the case of $i=k$.
\begin{figure*}
    \begin{equation}
    \label{eq:partial-R}
    \frac{\partial R_k}{\partial x_{\text{p}}^{i,j}}=\frac{-p_ip_k\upsilon_{k,k}}{\ln2\left(\sum\limits_{k'\neq k}^Kp_{k'}\upsilon_{k',k}+N\sigma_{k}^2+p_k\upsilon_{k,k}\right)\left(\sum\limits_{k'\neq k}^Kp_{k'}\upsilon_{k',k}+N\sigma_{k}^2\right)}\times\frac{\partial \upsilon_{i,k}}{\partial x_{\text{p}}^{i,j}}.
\end{equation}
\hrulefill
\end{figure*}
\end{appendices}

\bibliographystyle{IEEEtran}
\bibliography{mybib}
\end{document}